\documentclass[12pt]{article}
\PassOptionsToPackage{counterclockwise}{rotating}
\usepackage{eurosym}
\usepackage{textcomp}
\usepackage{multirow}
\usepackage[hyperfootnotes=false]{hyperref}
\usepackage[all]{hypcap}
\usepackage{listings}
\usepackage{amsfonts}
\usepackage{amsthm}
\usepackage{appendix}
\usepackage{setspace}
\usepackage{color}
\usepackage[margin=1in]{geometry}
\usepackage[small,bf]{caption}
\usepackage{longtable}
\usepackage{caption}
\usepackage{subcaption}
\usepackage{appendix}
\usepackage{tikz}
\usepackage[counterclockwise]{rotating}
\usepackage{ctable}
\usetikzlibrary{automata, positioning, arrows}
\usepackage{ifthen,graphics,graphicx,lscape,amsmath,dcolumn}
\usepackage{fancybox,epsf,latexsym,amssymb}
\usepackage[latin5]{inputenc}
\usepackage[all]{hypcap}
\usepackage{bbold}
\usepackage{verbatim}
\usepackage{epstopdf}
\usepackage{diagbox}
\usepackage{dcolumn}
\usepackage{bookmark}
\usepackage{cleveref}

\usepackage[longnamesfirst, authoryear, round]{natbib}

\theoremstyle{plain}
\providecommand{\U}[1]{\protect\rule{.1in}{.1in}}

\newtheorem*{theorem*}{Theorem}

\usepackage{collcell}
\usepackage{hhline}
\usepackage{pgf}
\usepackage{multirow}

\def\colorModel{hsb} 

\newcommand\ColCell[1]{
	\pgfmathparse{#1<50?1:0}  
	\ifnum\pgfmathresult=0\relax\color{white}\fi
	\pgfmathsetmacro\compA{0}      
	\pgfmathsetmacro\compB{#1/100} 
	\pgfmathsetmacro\compC{1}      
	\edef\x{\noexpand\centering\noexpand\cellcolor[\colorModel]{\compA,\compB,\compC}}\x #1
}
\newcolumntype{E}{>{\collectcell\ColCell}m{0.4cm}<{\endcollectcell}}  

\usepackage{siunitx}
\usepackage{xcolor}
\newcolumntype{R}{S[color=blue,negative-color = red]}

\definecolor{ChadDarkBlue}{rgb}{.1,0,.2}
\definecolor{ChadBlue}{rgb}{.1,.1,.5}
\definecolor{ChadRoyal}{rgb}{.2,.2,.8}
\definecolor{ChadGreen}{rgb}{0,.4,0}    
\definecolor{ChadRed}{rgb}{.8,.1,.2}    

\hypersetup{
	colorlinks=true,        
	linkcolor=ChadRed,
	urlcolor=ChadGreen,
	citecolor=ChadRed,
	pdftitle={CollateralValue},
	pdfauthor={ },
	pdfstartview={FitV},
	pdfpagemode={UseNone},
	pdfnewwindow=true,      
}

\begin{document}
\linespread{1.25}

\title{Disentangling Liquidity-Constrained from Strategic Default: A Deep Learning Approach}

\author{{
\begin{tabular}[t]{c}
Arka Prava Bandyopadhyay \thanks{%
Bandyopadhyay, ab3985@columbia.edu, Adjunct Assistant Professor, Columbia University. I thank Yildiray Yildirim, Johannes Stroebel, Lilia Maliar, Arpit Gupta, Agostino Capponi, Kay Giesecke, Christopher Palmer, Brent Ambrose, Sumit Agarwal, Sasha Indarte, Rawley Heimer, Anthony DeFusco, Natalya Schenck, Dominik Rehse, Linda Allen and Stefania Albanesi for helpful comments, and conference participants at ASSA 2020, FMA 2020, the Bloomberg BBQ 2020, BlackRock FMG 2020, the Baruch College Annual Real Estate Symposium 2020, and the University of Florida Machine Learning Conference 2022. All remaining errors are my own.}
\end{tabular}
}}

\date{\today}

\maketitle
\thispagestyle{empty}
\bigskip

\begin{abstract}
\noindent I use a deep neural network (DNN) on a proprietary panel of commercial mortgages spanning 1998--2021 to separate liquidity-constrained default from strategic default -- two competing, observationally similar explanations for delinquency that are difficult to disentangle with standard parametric models. Conditional on delinquency, property income (net operating income, or NOI) is a more important determinant of the default rate than leverage (loan-to-value, or LTV), despite LTV's status as the workhorse variable in the commercial mortgage default literature. I recover borrower motive from otherwise observationally equivalent delinquency states through a joint analysis of default rates across NOI and LTV buckets, and show that the resulting thresholds are stable across the 2008 financial crisis and the 2020--2021 COVID-19 shock to commercial real estate. Because commercial borrowers are firms rather than households, and because responsibility for a missed payment is diffused across an organization rather than borne by a single decision-maker, the moral and reputational frictions that discourage strategic default in the residential setting are attenuated -- making commercial mortgages a setting in which strategic behavior should be comparatively easier to detect, and comparatively costly to ignore for investors, servicers, and regulators.
\end{abstract}



\clearpage

\section{Introduction}
\label{sec:intro}

A commercial mortgage borrower who stops paying may be doing one of two very different things. She may be unable to pay: the property's net operating income (NOI) has fallen short of the debt obligation, and delinquency is a mechanical consequence of a cash-flow shortfall. Or she may be unwilling to pay: the property's contemporaneous value has fallen far enough below the outstanding balance that continuing to service the loan no longer makes financial sense, and delinquency is a deliberate, calculated decision -- what the literature terms \emph{strategic default} (\citet{FosterVanOrder1984}; \citet{Riddiough1991}). These two stories generate the same observed outcome -- a missed payment -- and are therefore extremely difficult to tell apart using only delinquency status as an outcome variable. Yet distinguishing them matters enormously for how investors price commercial mortgage-backed securities (CMBS), how servicers decide whether to modify or foreclose, and how regulators design countercyclical capital and forbearance policy.

This paper uses a large, proprietary loan-level panel of commercial mortgages from Trepp -- the leading data provider for the CMBS market -- together with a deep neural network (DNN), to separate these two motives. The identification strategy exploits a simple observation. If default were purely a function of liquidity, the default rate would be roughly flat, or even decreasing, in NOI: better cash flow should mean fewer missed payments. If default instead reflects a strategic choice conditioned on negative or eroding equity, the default rate can \emph{increase} in NOI within some LTV ranges -- borrowers who have positive cash flow but severely impaired collateral value may nonetheless choose to walk away, since the property no longer justifies continued debt service relative to its outside option. I show that both patterns exist in the data, and that which one dominates depends systematically on where a loan sits in the joint distribution of LTV and NOI. The DNN's chief empirical advantage here is not classification accuracy for its own sake, but its ability to recover these regime-dependent, non-monotonic relationships and the LTV/NOI thresholds that separate them, without the analyst having to specify a functional form or an interaction structure in advance.

The empirical setting -- commercial rather than residential mortgages -- is central to the identification. \citet{Ganong} find that only about 6\% of underwater residential defaults are caused exclusively by negative equity -- an order of magnitude lower than the literature had previously assumed -- in part because households face substantial reputational, psychological, and non-recourse-law frictions that discourage walking away even when underwater (\citet{NBERw15145}). Commercial mortgage borrowers face a very different incentive structure. They are firms, not families; the decision to miss a payment is made by an agent (an asset manager, a property-level LLC) on behalf of a diffuse set of principals, and the reputational cost of a single loan's delinquency is not borne personally by one individual in the way a foreclosure is borne by a homeowner. Institutional commercial borrowers are also professional investors who evaluate a levered real asset largely -- if not exclusively -- on its risk-adjusted cash-flow economics. Where the residential default literature has to work hard to isolate strategic behavior from life events and behavioral frictions, the commercial mortgage setting should make such behavior comparatively easier to detect, and its absence from most parametric commercial mortgage models comparatively more consequential.

The paper's substantive finding runs against the grain of the existing commercial mortgage literature, which has generally treated LTV as the primary -- sometimes sole -- driver of default, motivated by the put-option structure of non-recourse debt (\citet{Ambrose2001OptimalPE}; \citet{Ambrose2012}). I do not dispute that LTV matters. I show, however, that once the model is allowed to capture non-linearities and interactions flexibly, NOI is a more powerful and more robust predictor of delinquency than LTV, and that this ranking survives (i) leave-one-variable-out robustness checks on other contractual features (prepayment penalty, balloon indicator, year-month fixed effects), (ii) an out-of-sample stress test trained only on pre-2008 data and tested through the financial crisis, and (iii) an analogous test trained on pre-pandemic data and tested through the 2020--2021 COVID-19 shock to commercial real estate. Parametric alternatives -- multinomial logistic regression, Lasso, and Ridge -- and even flexible tree-based methods such as random forests and gradient boosting, do not reproduce this ranking reliably; only the DNN's higher-order interaction structure recovers it consistently across sub-periods.

This paper contributes to the real estate finance literature in three ways. First, it provides new loan-level evidence on the relative importance of liquidity versus leverage in commercial mortgage default -- a long-standing question in the literature dating to \citet{FosterVanOrder1984} and \citet{Riddiough1991} -- using a data set an order of magnitude larger than what is typical in this literature (91,767 loans, 9.6 million loan-month observations, 1998--2021) and a full crisis-plus-pandemic sample period. Second, it develops a bivariate LTV--NOI diagnostic that identifies observable thresholds along both dimensions at which borrower behavior shifts between liquidity-constrained and strategic patterns, giving practitioners and regulators a concrete, data-driven rule of thumb rather than a purely theoretical distinction. Third, it demonstrates -- with explicit robustness and stability checks against simpler machine learning benchmarks -- that flexible non-parametric methods can serve as a credible identification device in settings where standard quasi-experimental tools (difference-in-differences, regression discontinuity, instrumental variables) are difficult to implement because of the absence of a clean, exogenous shock at the loan level.

The remainder of the paper proceeds as follows. Section \ref{sec:institutional} discusses the institutional differences between commercial and residential mortgage default that motivate the empirical strategy. Section \ref{sec:lit} situates the paper within the strategic default and commercial mortgage literatures. Section \ref{sec:Theory} develops a simple discrete-time model of the borrower's default-timing decision that motivates the choice of covariates. Section \ref{sec:data} describes the Trepp data. Section \ref{sec:models} lays out the empirical models, from parametric benchmarks through the DNN. Section \ref{sec:results} presents the results, including the LTV--NOI threshold analysis, variable-importance comparisons across models, and the financial-crisis and COVID-19 robustness checks. Section \ref{sec:conclusion} concludes with implications for CMBS investors, servicers, and policy.

\section{Institutional Background: Commercial Versus Residential Mortgage Default}
\label{sec:institutional}

The residential real estate boom of 2004--2008 and its subsequent collapse have generated an enormous scholarly and policy literature. The parallel run-up in commercial real estate (CRE) valuations over the same period has received comparatively little attention, despite outstanding CMBS balances that have at times approached a trillion dollars. Since securitization became the dominant financing channel for commercial mortgages in the early 2000s, competition among increasingly aggressive ``originate-to-securitize'' lenders is widely believed to have eroded underwriting standards in the CRE market over time (\citet{WachterLevitin}). Despite some overlap in the multifamily segment, commercial and residential real estate are economically distinct markets that have attracted correspondingly different forms of government intervention (for example, GSE support is largely confined to residential and multifamily lending).

The central institutional difference this paper exploits is that commercial mortgage default depends not just on the value of the collateral (borrower equity) but also, independently, on the property's ability to generate income (property liquidity). Non-recourse commercial and residential debt both carry an implicit put option: the borrower can extinguish the debt by surrendering the collateral, with the property's value as the effective strike price. This put-option logic is the standard theoretical justification for treating LTV as the primary driver of default in the prior literature (\citet{Ambrose2012}; \citet{Ambrose2001OptimalPE}). But a rational borrower generally will not exercise this option while net property income is sufficient to service scheduled debt, even when book equity is negative -- and, conversely, a borrower with severely impaired equity may delay exercise if positive cash flow continues to accrue, preserving the option value of defaulting later under worse conditions. A complete model of commercial mortgage default therefore needs both value and income as candidate triggers, which is precisely the joint LTV--NOI structure this paper exploits empirically.

A second institutional difference concerns amortization. Unlike most residential mortgages, which are fully amortizing, commercial mortgages are typically partially amortizing, with a 7--12 year term against a 25--30 year amortization schedule -- meaning a balloon payment falls due at maturity. Borrowers ordinarily fund this balloon payment by refinancing, which can become difficult -- and hence strategically fraught -- exactly when interest rates rise or underwriting standards tighten, even for an otherwise well-performing loan. I observe a marked increase in loans that appear to linger deliberately in the 90--120 day delinquency bucket after the 2008 financial crisis, consistent with borrowers using this status as a way of avoiding formal foreclosure and real-estate-owned (REO) classification while still not curing the loan.

A third difference concerns the borrower's identity and the resulting information environment. The lender must price a loan against an unobserved distribution of borrower ``types'' at origination -- generating classic adverse selection -- and cannot perfectly monitor the borrower's ongoing willingness, as opposed to ability, to pay after origination, generating moral hazard. Because commercial borrowers compete for lessees in local, segmented property markets, there is additional search-friction-driven uncertainty in the NOI a given property will generate through its lease-up and renewal cycle. I argue that some borrowers use strategic delinquency -- particularly the extended 90--120 day bucket described above -- as a kind of self-provided insurance against the risk premium embedded in the mortgage rate charged at origination.

These three institutional features -- income as an independent default trigger, balloon-driven refinancing risk, and diffused, firm-level decision-making -- motivate treating commercial mortgage credit risk as jointly determined by market risk, credit risk, prepayment risk (\citet{ChristopoulosJarrowYildirim2008}), and liquidity risk (\citet{AmbroseSanders2003}), rather than by leverage alone. The remainder of the paper develops this argument empirically.

\section{Related Literature}
\label{sec:lit}

\subsection{Strategic Default in Residential and Commercial Real Estate}

The theoretical roots of this paper's central distinction trace to two competing accounts of mortgage default. \citet{Riddiough1991} models default as triggered by an adverse life event that reduces borrower cash flow. \citet{FosterVanOrder1984} instead model default as an option exercised when negative equity makes continued ownership -- viewed as a financial asset -- unattractive regardless of the borrower's personal financial circumstances, giving rise to the ``strategic'' or ``ruthless'' default terminology. \citet{FooteGerardiWillen2008} argue that both channels typically operate jointly: negative equity is a necessary but not sufficient condition for default, which also requires a triggering adverse income shock, rather than either alone. \citet{FooteGerardi2009} extend this skepticism of a single-trigger view of default to the design of foreclosure-prevention policy.

In residential markets, \citet{NBERw15145} use survey data to estimate that roughly 26\% of existing defaults are strategic, and that moral and social considerations -- not just the size of the equity shortfall -- are important predictors of stated willingness to default; essentially no respondent reports willingness to default with an equity shortfall below 10\% of home value. \citet{Ganong} similarly find a low share of default caused exclusively by negative equity (about 6\%) once life-event and adverse cash-flow triggers are properly accounted for -- with the large majority of underwater defaults instead classified as ``double-trigger'' events requiring both negative equity and a negative life event -- underscoring how difficult strategic behavior is to isolate even with rich household-level data and a clear negative-equity definition.

Commercial borrowers differ from households along exactly the dimensions that the residential literature identifies as frictions against strategic default. They are firms rather than individuals, so the blame for a missed payment is organizationally diffused rather than personally borne; they are professional, repeat market participants whose delinquency decisions are plausibly closer to a profit-and-loss calculation than to a household's consumption-smoothing problem; and non-recourse and foreclosure law, while still relevant, interacts with corporate rather than personal liability. This paper's empirical strategy -- examining default rates jointly across LTV and NOI buckets -- is designed to exploit exactly this setting: a context in which the theoretical case for strategic behavior is strongest and the institutional frictions that dampen it in the household literature are comparatively weak.

\subsection{Commercial Mortgage and CMBS Default Literature}

The empirical CMBS and commercial mortgage literature has historically emphasized a set of loan- and market-level covariates: creditworthiness and free cash flow at the entity level, current leverage, loan age, interest rates, and CMBS index levels (\citet{FurstenbergGeorge1969}; \citet{CurleyGuttentag1974}; \citet{CampbellDietrich1983}; \citet{CunninghamCapone1990}; \citet{Deng1999}). \citet{AnDengGabriel2009} and \citet{AmbroseSanders2003} study CMBS loan pricing and the competing risks of default and prepayment, respectively, and both use variable sets close to the one adopted here. \citet{ChildsOttRiddiough1996} study the pricing of multi-class CMBS structures and the interaction between pool diversification and tranche-level value, motivating this paper's inclusion of deal-level structural variables (number of properties per deal). \citet{TitmanTomaidisTsyplakov2005} study the determinants of credit spreads on commercial mortgages, and \citet{YildirimKauKeenan2009} and \citet{Yildirim2008} develop structural and survival-based approaches, respectively, to estimating CMBS default probabilities, the latter using a frailty/mixture-model approach to separate long-term survivorship from default timing -- a different but complementary strategy to the one used here for separating default \emph{motive}.

This paper's findings differ from -- and in some respects run counter to -- a body of work that has treated loan-to-value as the primary determinant of commercial mortgage borrower behavior (\citet{CampbellDietrich1983}; \citet{CunninghamCapone1990}; \citet{Deng1999}; \citet{ElulHunt2010}; \citet{FooteGerardi2009}). I do not claim LTV is unimportant; I show that it is dominated, in both predictive and economic terms, by property income once the empirical model is allowed to capture the non-linear interactions between the two. \citet{LimboLinda2015} documents ``limbo'' loans that linger in extended delinquency without resolving to either cure or liquidation in the residential context; I document an analogous, and in some respects more pronounced, pattern in commercial loans around the 90--120 day and beyond-120-day delinquency buckets, particularly following the 2008 crisis.

\section{A Simple Model of Default Timing}
\label{sec:Theory}

This section develops a spare, discrete-time model of the commercial borrower's default-timing decision, in the spirit of \citet{NBERw15145} and \citet{Bajari2008}, to motivate the empirical covariates used below and to fix intuition for the LTV--NOI threshold structure documented in Section \ref{sec:results}.

Consider a borrower who, at time $t$, holds a property worth $A_t$ against a balloon payment $B_T$ due at loan maturity $T$. On purely financial grounds the borrower will not default so long as $A_t > B_T$. But the decision to default strategically also weighs considerations beyond the immediate financial gain or loss: by not defaulting today, the borrower preserves the option to default later, under potentially worse conditions -- for example, at a point closer to maturity when refinancing has become difficult, or after an adverse shock to occupancy. This intertemporal trade-off is further shaped by the timing of appraisal reductions and by whether the master or special servicer has ceased advancing principal and interest, both of which affect the near-term consequences of remaining current. Because lenders are typically close to NPV-neutral and default represents a deadweight loss for them, defaulting borrowers also face a higher future cost of credit through differential credit rationing. Let $K_t$ denote the net benefit, or opportunity cost of cash, from not defaulting at $t$. A rational borrower does not default at $t$ if
\begin{equation}
    A_t - B_T + K_t > 0.
\end{equation}

If no balloon payment is imminent, the decision is more complex still, because the borrower trades off defaulting today against the option to default tomorrow -- and that future option is itself conditional on the borrower's future ability to service the loan, captured empirically by the debt-service coverage ratio (DSCR), which is in turn highly correlated with occupancy and lessee cash flow. If the property sits vacant or the lessee stops paying, the borrower is likely to default in the following period regardless, extinguishing the value of waiting. Let $V_T = A_T - B_T + K_T$ denote the terminal value of not defaulting at maturity $T$. The value of not defaulting one period earlier is
\begin{equation}
    V_{T-1} = a_{T-1} - m_{T-1} - B_T + K_{T-1} + (1 - \pi_{T-1})\, \mathbb{E}\!\left[\max(V_T, 0)\right],
\end{equation}
where $a$ is the monetary value of the cash flow and service flow the property generates between $T-1$ and $T$, $m$ is the scheduled and unscheduled mortgage payment due over the same interval, and $\pi_{T-1}$ is the probability the property is vacant (has no paying lessee) over that interval. By backward induction, the value of not defaulting at a generic date $t$ is
\begin{equation}
    V_t = a_t - m_t - B_T + K_t + (1-\pi_t)\, \mathbb{E}\!\left[\max(V_{t+1}, 0)\right].
\end{equation}

The decision to default at a generic time $t$ can then be summarized as
\begin{equation}
    \text{Default}_t = F(A_t - B_T,\; a_t,\; m_t,\; \pi_t,\; K_t).
\end{equation}

The functional form of $F(\cdot)$ is, in general, extremely difficult to pin down analytically, and even a piecewise, locally linearized characterization via the Implicit Function Theorem would require strong regularity conditions on $F^{-1}$ with respect to each argument. Section \ref{sec:results} shows that these regularity conditions plainly fail in the data: the relationship between the default rate and each of these state variables is highly non-linear and, in the case of NOI, non-monotonic within certain LTV ranges. This motivates the paper's empirical strategy directly. LTV is a (noisy) empirical proxy for the equity shortfall $A_t - B_T$; NOI is a proxy for the cash-flow-and-service-flow term $a_t$; DSCR proxies for $a_t/m_t$; occupancy proxies for the borrower's belief about vacancy risk, $1-\pi_t$; and the non-monetary benefit $K_t$ of continued payment is, by its nature, largely unobservable and must instead be inferred indirectly, from the shape of the estimated relationship between default and the observable state variables. This is further complicated in practice by the uneven application of recourse law across states, which requires correspondingly strong state-level bankruptcy law to be effective in reinforcing recourse provisions. Given this, rather than impose a parametric form on $F(\cdot)$, Section \ref{sec:models} adopts a flexible non-parametric approach -- the DNN -- designed to recover $F(\cdot)$'s shape, including its thresholds and regime changes, directly from the data.

\section{Data}
\label{sec:data}

I use a monthly, loan-level panel of 91,767 U.S. commercial mortgages from January 2000 through September 2016 (with an extension through September 2021 used for the COVID-19 robustness analysis in Section \ref{sec:results}), obtained from Trepp, the leading provider of analytics and data to the CMBS, commercial mortgage finance, and banking industries. Trepp's underlying database covers more than 1,800 deals and over 100,000 loans supporting close to \$800 billion in securities across North American, European, and Asian CMBS, as well as CRE-backed CDOs; I restrict the sample to CMBS loans, since NOI -- the paper's central variable -- is reported only for income-producing properties financed through CMBS structures, and exclude balance-sheet CRE loans on this basis. After cleaning, the panel contains 9,617,333 loan-month observations of continuous variables; summary statistics are reported in Table \ref{Summary Statistics}.

I follow the covariate set established in the prior CMBS literature (\citet{AnDengGabriel2009}; \citet{AmbroseSanders2003}), which typically includes the log of original balance, LTV, amortization term, time to maturity, lockout and lockout expiration, corporate bond credit spreads (\citet{TitmanTomaidisTsyplakov2005}), the Treasury yield curve, the mortgage-Treasury rate spread, and region and seasonal fixed effects. I ultimately retain: loan-to-value (LTV), occupancy rate, tranche-level loan-to-value and weighted average coupon, the annualized gross note rate, outstanding scheduled principal balance, the most recently reported NOI, the outstanding legal principal balance net of defeasance, the securitized loan balance pledged to the trust, the most recent appraised value, scheduled and unscheduled principal payments, prepayment penalties, and net proceeds, expenses, and realized losses associated with liquidation. I include property age and its square as controls for the non-linear relationship between loan seasoning and delinquency, and construct a ``time to maturity'' variable specifically to detect strategic behavior concentrated near loan maturity.

Deal- and property-level controls include loan vintage (to capture the effect of time-varying underwriting standards), state fixed effects (metro, county, and zip-level identifiers have severe missingness and are not used), an eight-category property-type classification, a fixed/floating rate indicator, and the number of properties per deal. I assign a maturity-proximity dummy to loans within three months of their scheduled maturity date to flag exposure to refinancing and balloon risk. I augment the loan-level panel with the MIT Commercial Real Estate Price Index, NCREIF regional property value indices, state-level quarterly GDP (converted to a monthly frequency), state-level monthly unemployment, and Treasury rates of multiple maturities.

All continuous covariates are min-max normalized to $[0,1]$ prior to estimation, $x \mapsto \frac{x - \min(x)}{\max(x) - \min(x)}$, rather than standardized to a common variance, in order to preserve each variable's underlying skewness and other distributional features -- information that a standard $z$-score transformation would discard -- while still placing all covariates on a common, model-friendly scale.

\begin{table}[!htbp]
\centering
  \caption{Summary statistics for the cleaned panel of 9,617,333 loan-month observations of continuous variables. \textbf{M} denotes millions and \textbf{B} denotes billions. Categorical variables are one-hot encoded prior to estimation. Because variables differ substantially in scale and skewness, and because model output is a probability on $[0,1]$, all continuous variables are subsequently min-max mapped to $[0,1]$.}
  \label{Summary Statistics}
\begin{tabular}{@{\extracolsep{-5pt}}lccccccc}
\\[-1.8ex]\hline
\hline \\[-1.8ex]
Statistic & \multicolumn{1}{c}{N} & \multicolumn{1}{c}{Min} & \multicolumn{1}{c}{Pctl(25)} & \multicolumn{1}{c}{Median} & \multicolumn{1}{c}{Mean} & \multicolumn{1}{c}{Pctl(75)} & \multicolumn{1}{c}{Max} \\
\hline \\[-1.8ex]
beginbal & 9,617,333 & 0 & 1,838,146 & 4,036,697 & 7,882,727 & 9,100,000 & 99,990,043 \\
orig\_bal & 9,617,333 & 0 & 1.92M & 4.10M & 9.31M & 9.14M & 1.68B \\
rate & 9,617,333 & 0 & 6 & 6 & 6 & 7 & 9 \\
Sched\_princip & 9,617,333 & 0 & 1,060 & 4,235 & 15,665 & 9,838 & 430M \\
Unsched\_prin & 9,617,333 & 0 & 0 & 0 & 50,300 & 0 & 1.50B \\
balance\_act & 9,617,333 & 0 & 1,774,511 & 3,980,981 & 7,797,502 & 9,009,824 & 99,999,000 \\
payment & 9,617,333 & 0 & 15,000 & 28,600 & 65,900 & 57,800 & 675M \\
pppenalties & 9,617,333 & 0 & 0 & 0 & 420 & 0 & 29,477,125 \\
liqproceeds & 9,617,333 & 0 & 0 & 0 & 17,900 & 0 & 2.56B \\
realizedloss & 9,617,333 & 0 & 0 & 0 & 5,260 & 0 & 204M \\
liqexpense & 9,617,333 & 0 & 0 & 0 & 3,200 & 0 & 1.06B \\
numprop & 9,617,333 & 1 & 1 & 1 & 1 & 1 & 225 \\
Appraisal\_Reduc & 9,617,333 & 0 & 0 & 0 & 132,000 & 0 & 391M \\
SecurLTV & 9,617,333 & 0 & 63 & 71 & 67 & 76 & 150 \\
Face & 9,617,333 & 0 & 6 & 6 & 6 & 7 & 9 \\
NOI & 9,617,333 & 0 & 268,000 & 527,000 & 1.22M & 1.10M & 1.09B \\
LTV & 9,617,333 & 0 & 63 & 71 & 69 & 77 & 150 \\
AppValue & 9,617,333 & 1,620 & 3.46M & 6.80M & 17.4M & 14.5M & 48.1B \\
OccRate & 9,617,333 & 0.0 & 0.89 & 0.96 & 0.92 & 1.0 & 1.4 \\
Basis & 9,617,333 & 0.0 & 2.0 & 2.0 & 1.7 & 2.0 & 4.0 \\
Unemp & 9,617,333 & 0.019 & 0.049 & 0.061 & 0.066 & 0.080 & 0.154 \\
GDP & 9,617,333 & -0.020 & 0.008 & 0.016 & 0.013 & 0.048 & 0.0174 \\
2YrTr & 9,617,333 & 0.002 & 0.006 & 0.010 & 0.019 & 0.031 & 0.061 \\
10YrTr & 9,617,333 & 0.015 & 0.024 & 0.036 & 0.034 & 0.043 & 0.063 \\
NAREIT & 9,617,333 & -0.45 & -0.05 & 0.01 & -0.01 & 0.04 & 0.38 \\
MIT.Liq & 9,617,333 & 0.017 & 0.022 & 0.036 & 0.025 & 0.105 & 0.184 \\
time\_to\_maturity & 9,617,333 & 0 & 35 & 66 & 80 & 100 & 765 \\
age & 9,617,333 & 0 & 31 & 60 & 63 & 90 & 409 \\
age$^2$ & 9,617,333 & 0 & 961 & 3600 & 5393 & 8100 & 167281 \\
\hline \\[-1.8ex]
\end{tabular}
\end{table}

\section{Empirical Models}
\label{sec:models}

I classify loans into seven delinquency states, moving from current to fully resolved: \textbf{W0\_30D} (current, or within the grace period), \textbf{W30\_60D} and \textbf{W60\_90D} (late payment of 30--60 and 60--90 days, respectively), \textbf{W90\_120D} (90--120 days delinquent, the paper's operational default threshold), \textbf{B120D} (beyond 120 days, encompassing liquidation and final resolution), and two additional maturity-proximate states, \textbf{PrfMatBal} (performing, mature, with a balloon payment due) and \textbf{NPrfMatBal} (non-performing, mature, with a balloon payment due), which I treat as distinct states because loans approaching maturity are disproportionately exposed to strategic refinancing risk even while nominally current. I estimate models in increasing order of flexibility, culminating in the DNN, and evaluate all models on both in-sample cross-validation error and true out-of-sample prediction on a later holdout period.

\subsection{Parametric Benchmarks}

I begin with a set of standard parametric classifiers -- ordinal logistic regression, multinomial logistic regression, and Lasso- and Ridge-regularized logistic regression -- to establish a baseline and to illustrate why these models are ill-suited to the joint classification problem at hand. The ordinal model respects the natural ordering of delinquency severity but, by construction, cannot represent transitions from a worse to a better delinquency state. The multinomial model instead imposes the Independence of Irrelevant Alternatives (IIA) assumption, which is directly violated here: introducing an intermediate delinquency state (say, 30--60 days) changes the relative odds of a borrower choosing to be ``current'' versus ``90--120 days delinquent,'' since some borrowers will now choose the intermediate state opportunistically rather than either extreme -- exactly the kind of strategic substitution this paper is designed to detect. Appendix \ref{Appendix: Multinomial Logistic Model} develops this point formally via a finite-state-automaton representation of feasible delinquency transitions.

\subsection{Tree-Based Machine Learning Benchmarks}

I next estimate a distributed random forest (DRF) and a gradient boosting machine (GBM), both of which allow for non-linearity and interaction effects without a pre-specified functional form. Variable importance for the DRF is computed via permutation importance -- the degradation in out-of-bag prediction accuracy from randomly shuffling each covariate, averaged across trees -- following \citet{Khandani}. The GBM sums importance contributions across boosting iterations. As documented in Section \ref{sec:results}, both models improve substantially on the parametric benchmarks in classification accuracy, but neither reliably ranks NOI above LTV, and the GBM's importance ranking in particular proves unstable out-of-sample and during the financial crisis, consistent with its status as a greedy, locally-optimizing algorithm.

\subsection{Deep Neural Network}
\label{sec:dnn}

The paper's primary model is a multi-layer feedforward DNN, trained with stochastic gradient descent via back-propagation, using asynchronous multi-threaded model averaging across compute nodes. I address class imbalance in the delinquency outcome using the Synthetic Minority Oversampling Technique (SMOTE; \citet{DBLP:journals/jair/ChawlaBHK02}), and tune both optimizer- and model-specific hyperparameters via random grid search (full specification in Appendix \ref{sec:Hyperparameter}). I quantify each covariate's marginal contribution using Gedeon's connection-weight-based variable importance method (\citet{Goh1995BackpropagationNN}, described further in the appendix), which -- unlike model-specific coefficients -- is comparable across all seven models and does not require the network output to be re-scaled, since inputs are already normalized to $[0,1]$.

The economic case for the DNN, rather than a richer parametric model, rests on two properties. First, with sufficient hidden units, a feedforward network can approximate any continuous function on a closed, bounded set arbitrarily well (\citet{Hornik1991}), including whatever unknown, non-linear interaction structure characterizes $F(\cdot)$ in Section \ref{sec:Theory} -- without the analyst first having to guess that structure. Second, adding depth rather than width lets the network represent functions of a given complexity with exponentially fewer parameters than a shallow network would require (\citet{MontuferPAscanuChoBengio2014}), which matters directly here given the size of the covariate space (33 loan-, deal-, and macro-level variables) and the sample (9.6 million loan-months). Because this approximation capacity comes at the cost of interpretability, Section \ref{sec:results} places considerable weight on cross-model and cross-period stability of the resulting variable-importance rankings, rather than treating in-sample fit alone as evidence that the DNN has recovered an economically meaningful relationship.

\section{Results}
\label{sec:results}

\subsection{Classification Performance}

\begin{table}[!htbp]
\centering
\caption{I report the Cross-Validation Training Errors of Models across Delinquency Classes with sample ranging from Jan 2000 - June 2012. The different models Ordinal, Mult (Multinomial Logistic), Lasso, Ridge, DRF (Distributed Random Forest), GBM (Gradient Boosting Machine), DNN (Deep Neural Network) and the different delinquency classes are W0\_30D (Within 30 days), W30\_60D (Within 30-60 days), W60\_90D (Within 60-90 days), W90\_120D (Within 90-120 days), PrfMatBal (Performing Mature Balloon), NPrfMatBal (Non-Performing Mature Balloon), B120D (Beyond 120 days). From W90\_120D and worse delinquency classes, the misclassification error rates decrease with increasing model flexibility and DNN performs best overall.}
\label{tab:Train}
\begin{tabular}{rrrrrrrrrl}
\hline
               & \textbf{Ordinal} & \textbf{Mult} & \textbf{Lasso} & \textbf{Ridge} & \textbf{DRF} & \textbf{GBM} & \textbf{DNN} \\ 
\hline
\textbf{W0\_30D} & 1\% & 1\% & 1\% & 1\% & 1\% & 1\% & 0\% \\ 
\textbf{W30\_60D} & \textbf{61\%} & 100\% & 100\% & 100\% & \textbf{98\%} & 100\% & 100\%  \\ 
\textbf{W60\_90D} & \textbf{63\%} & 100\% & 100\% & 100\% & \textbf{98\%} & 100\% & 100\%  \\ 
\textbf{W90\_120D} & 36\% & 48\% & 48\% & 47\% & \textbf{21\%} & \textbf{27\%} & \textbf{21\%}  \\ 
\textbf{PrfMatBal} & 100\% & 100\% & 100\% & 100\% & \textbf{76\%} & \textbf{84\%} & \textbf{78\%}  \\ 
\textbf{NPrfMatBal} & 100\% & 67\% & 70\% & 74\% & \textbf{16\%} & \textbf{12\%} & \textbf{12\%}  \\ 
\textbf{B120D} & 100\% & 80\% & 87\% & 83\% & \textbf{22\%} & \textbf{21\%} & \textbf{20\%}  \\ 
\hline
\textbf{Totals} & 3\% & 2\% & 2\% & 2\% & 1\% & 1\% & 1\%  \\ 
\hline
\end{tabular}
\end{table}

\begin{table}[!htbp]
\centering
\caption{I report the Out-of-Sample Test Errors of Models across Delinquency Classes with sample ranging from Jul 2012 - June 2016. The different models Ordinal, Mult (Multinomial Logistic), Lasso. Ridge, DRF (Distributed Random Forest), GBM (Gradient Boosting Machine), DNN (Deep Neural Network) and the different delinquency classes are W0\_30D (Within 30 days), W30\_60D (Within 30-60 days), W60\_90D (Within 60-90 days), W90\_120D (Within 90-120 days), PrfMatBal (Performing Mature Balloon), NPrfMatBal (Non-Performing Mature Balloon), B120D (Beyond 120 days). From W90\_120D and worse delinquency classes, the misclassification error rates (in decimals) decrease with increasing model flexibility and DNN performs best overall.}
\label{tab:Test}
\begin{tabular}{rrrrrrrrrl}
\hline
 & \textbf{Ordinal} & \textbf{Multi} & \textbf{Lasso} & \textbf{Ridge} & \textbf{DRF} & \textbf{GBM} & \textbf{DNN} \\ 
\hline
\textbf{W0\_30D} & 1\% & 1\% & 1\% & 1\% & 1\% & 1\% & 1\% \\ 
\textbf{W30\_60D} & \textbf{89\%} & 100\% & 100\% & 100\% & 100\% & 100\% & 100\%  \\ 
\textbf{W60\_90D} & \textbf{59\%} & 100\% & 100\% & 100\% & 100\% & 100\% & 100\%  \\ 
\textbf{W90\_120D} & 20\% & 37\% & 37\% & 37\% & \textbf{14\%} & \textbf{15\%} & \textbf{17\%}  \\ 
\textbf{PrfMatBal} & 98\% & 100\% & 100\% & 100\% & \textbf{97\%} & \textbf{96\%} & \textbf{95\%}  \\ 
\textbf{NPrfMatBal} & 100\% & 42\% & 42\% & 47\%  & \textbf{19\%} & \textbf{13\%} & \textbf{13\%}  \\ 
\textbf{B120D} & 100\% & 100\% & 100\% & 100\% & \textbf{7\%} & \textbf{8\%} & \textbf{7\%}  \\ 
\hline
\textbf{Totals} & 98\% & 3\% & 3\% & 3\% & 3\% & 3\% & 3\%  \\ 
\hline
\end{tabular}
\end{table}

Table \ref{tab:Train} reports in-sample (training) misclassification rates across all seven models and all seven delinquency classes, using data from January 2000 through June 2012; Table \ref{tab:Test} reports genuinely out-of-sample rates on a July 2012--June 2016 holdout. All models classify the \textbf{W0\_30D} (current) class essentially perfectly, since it is the modal state. The parametric models -- multinomial logistic, Lasso, and Ridge -- misclassify the intermediate \textbf{W30\_60D} and \textbf{W60\_90D} states almost completely (rates near 100\%), consistent with the IIA violation discussed above: these models cannot distinguish a borrower who has strategically missed one payment from a borrower who is fully current. The ordinal model performs somewhat better on these intermediate classes precisely because it encodes the ordering of severity rather than treating states as independent, unordered alternatives. The tree-based models (DRF, GBM) and the DNN substantially outperform all parametric benchmarks on the more severe classes -- \textbf{W90\_120D}, \textbf{PrfMatBal}, \textbf{NPrfMatBal}, and \textbf{B120D} -- with out-of-sample misclassification rates for \textbf{B120D} falling from effectively 100\% under the parametric models to single digits under DRF, GBM, and the DNN. Prediction quality for all non-parametric models deteriorates modestly during the financial crisis window, as expected, but remains far superior to the parametric benchmarks throughout.

\subsection{The LTV--NOI Threshold Structure}

\begin{figure}
		\centering
		\includegraphics[width=250pt,height=250pt]{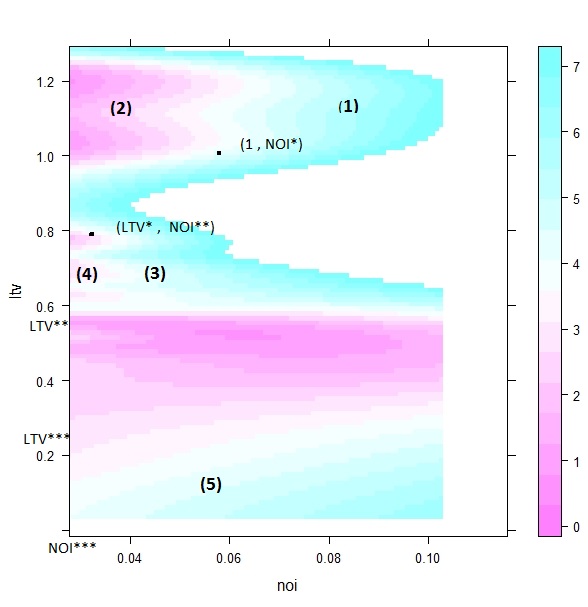}
		\caption{\textbf{Default Rate Vs NOI \& LTV}\\I list the possible combinations of LTV and NOI that can point to the incentives of default behavior.}
		\label{BivariateHeatmap_b}
		
	\end{figure} 

\begin{figure}[!htbp]
\centering
\includegraphics[width=3in]{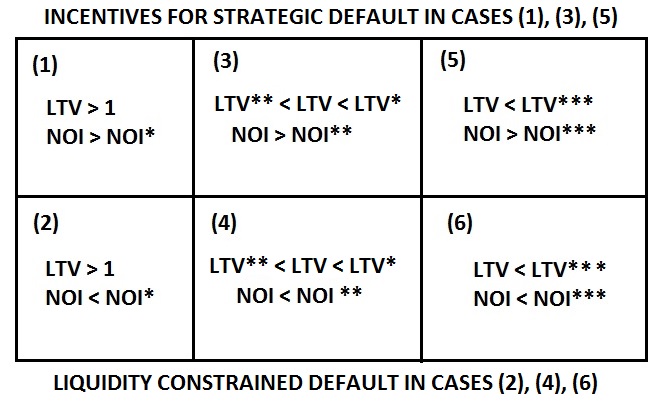}
\caption{I list the possible combinations of LTV and NOI that can \emph{disentangle} Liquidity-constrained Default and the incentives for Strategic Default behavior in this figure. }
\label{fig:Interaction of LTV and NOI}
\end{figure}

The paper's central empirical result is summarized by the bivariate relationship between the default rate and LTV and NOI jointly (Figure \ref{BivariateHeatmap_b}; Figure \ref{fig:Interaction of LTV and NOI}). When $LTV > 1$ -- the property is underwater and the borrower is, in a mechanical sense, insolvent -- the default rate \emph{increases} in NOI. This is the diagnostic signature of strategic default: among insolvent borrowers, those with more cash on hand are more, not less, likely to default, consistent with a borrower who can afford to pay but chooses not to because the collateral no longer justifies continued service. The DNN identifies a threshold $NOI^{*}$ (the sixth percentile of the Trepp NOI distribution) that separates this pattern from a more conventional liquidity-constrained pattern at lower LTV.

For $LTV^{*}(0.82) < LTV < 1$ -- solvent but highly levered -- default rates are elevated across nearly all borrowers, consistent with borrowers who are effectively passing NOI risk through to the lender in order to preserve remaining equity, anticipating the discrete jump to insolvency that a further appraisal reduction could trigger. For $LTV^{**}(0.60) < LTV < LTV^{*}(0.82)$ -- moderate leverage, no immediate negative-equity risk -- default instead begins at a much lower NOI threshold, $NOI^{**}$ (the fourth percentile), consistent with genuine liquidity-constrained default rather than strategic behavior. In the range $LTV^{***}(0.20) < LTV < LTV^{**}(0.60)$, default rates are low across the board: borrowers here face neither meaningful negative-equity risk nor binding liquidity constraints. Finally, for $LTV < LTV^{***}(0.20)$ -- loans close to maturity, where most of the original balance has been amortized or paid down -- default rates rise sharply again, consistent with the balloon-payment and refinancing-risk channel discussed in Section \ref{sec:institutional} rather than with either leverage or liquidity per se.

\begin{figure}[!htbp]
	\centering
	\begin{subfigure}{0.48\textwidth}
		\centering
		\includegraphics[width=1\linewidth,height=178pt]{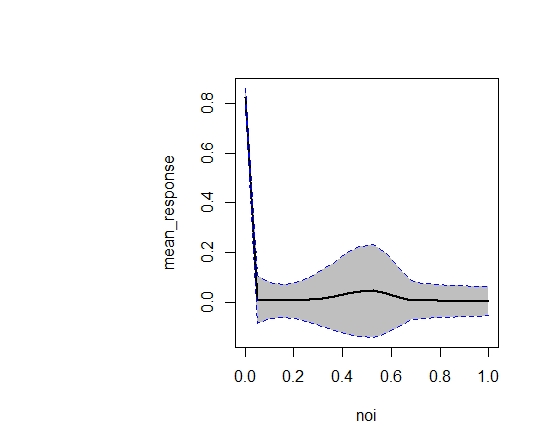}
		\caption{Default Rate Vs Net Operating Income}
		\label{PD_a}
		
	\end{subfigure}
	\begin{subfigure}{0.48\textwidth}
		\centering
		\includegraphics[width=1\linewidth,height=178pt]{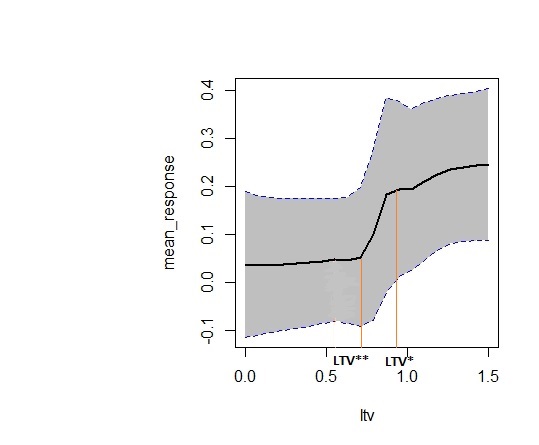}
		\caption{Default Rate Vs Loan-to-Value}
		\label{PD_b}
		
	\end{subfigure} \\
	%
	%
	%
	%
	%
	%
	
	\caption{Partial Dependence Plots for Predicted Default Rate}
	\label{PD_graphs}
\end{figure}

The DNN recovers these same thresholds from the univariate partial dependence plots for NOI and LTV separately (Figure \ref{PD_a}; Figure \ref{PD_b}), using all 33 covariates and capturing the full set of higher-order interactions and non-linearities among them -- something the bivariate heatmap, by construction, cannot do on its own. None of the parametric or tree-based alternatives recovers this threshold structure with comparable stability, as discussed next.

\subsection{Variable Importance: NOI Versus LTV}

\begin{figure}[!htbp]
\begin{subfigure}{0.3\textwidth}
  \centering
  \includegraphics[angle=90,width=.8\linewidth]{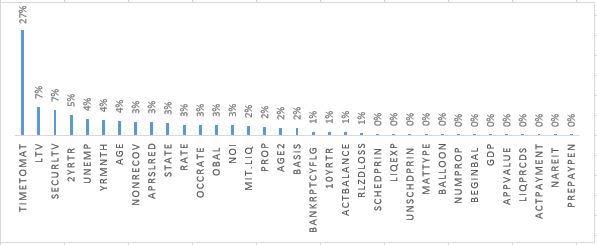}\quad
  \caption{Variable Importance: Lasso}
  \label{fig:Lasso}
\end{subfigure}%
\begin{subfigure}{0.3\textwidth}
  \centering
  \includegraphics[angle=90,width=.8\linewidth]{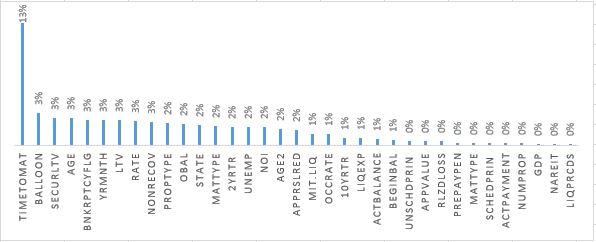}\quad
  \caption{Variable Importance: Ridge}
  \label{fig:Ridge}
\end{subfigure}
\begin{subfigure}{0.3\textwidth}\quad
  \centering
  \includegraphics[angle=90,width=.8\linewidth]{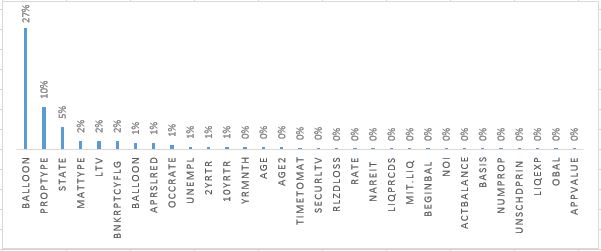}
  \caption{Variable Importance: Ordinal}
  \label{fig:Ordinal}
\end{subfigure}
\medskip
\begin{subfigure}{0.3\textwidth}
  \centering
  \includegraphics[angle=90,width=.8\linewidth]{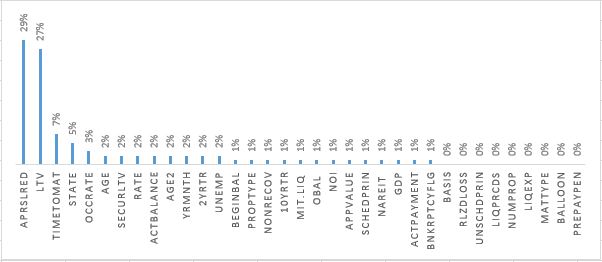}\quad
  \caption{Variable Importance: DRF}
  \label{fig:DRF}
\end{subfigure}
\begin{subfigure}{0.3\textwidth}
  \centering
  \includegraphics[angle=90,width=.8\linewidth]{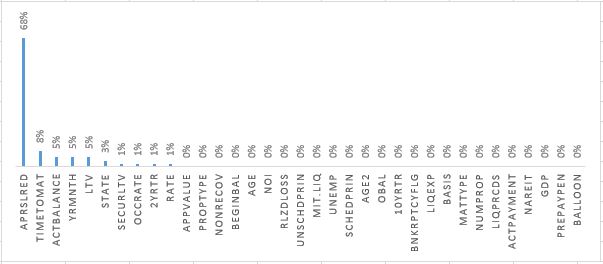}\quad
  \caption{Variable Importance: GBM}
  \label{fig:GBM}
\end{subfigure}
\begin{subfigure}{0.3\textwidth}
  \centering
  \includegraphics[angle=90,width=.8\linewidth]{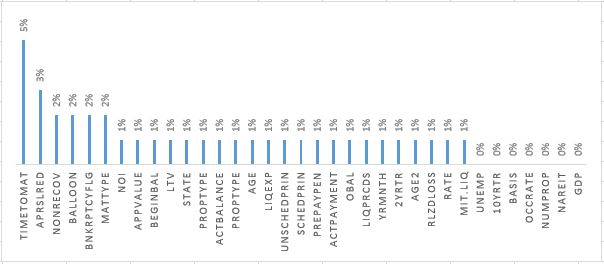}
  \caption{Variable Importance: DNN}
  \label{fig:DNN}
\end{subfigure}
\caption{Variable Importance for 6 models, namely, Lasso, Ridge, Ordinal, Distributed Random Forest (DRF), Gradient Boosting Machine (GBM), Deep Neural Network (DNN). Only the variables having more than 0.5\% marginal contribution are included in the variable importance visualizations for brevity.}
\label{fig:images}
\end{figure}

\begin{figure}[!htbp]
\begin{subfigure}{0.3\textwidth}
  \centering
  \includegraphics[angle=-90,width=.8\linewidth]{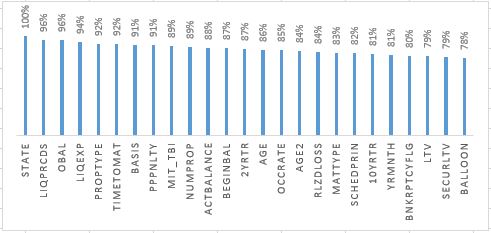}\quad
  \caption{VI without NOI}
  \label{fig:VIwithoutNOI}
\end{subfigure}%
\begin{subfigure}{0.3\textwidth}
  \centering
  \includegraphics[angle=-90,width=.9\linewidth]{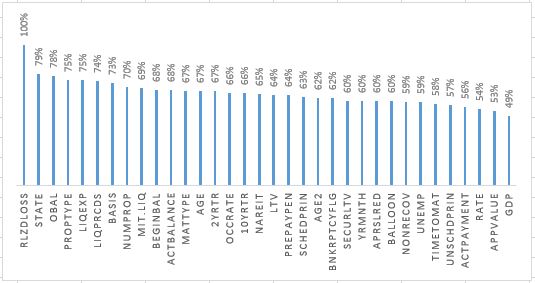}\quad
  \caption{VI without Year\_Month}
  \label{fig:VIwithoutYrMnth}
\end{subfigure}
\begin{subfigure}{0.3\textwidth}\quad
  \centering
  \includegraphics[angle=-90,width=.8\linewidth]{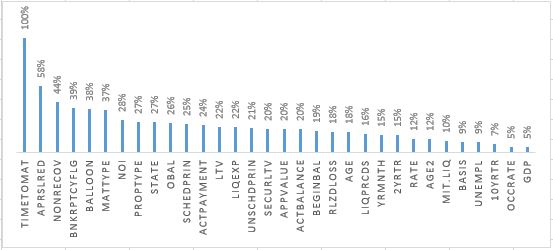}
  \caption{VI without PrePayPen}
  \label{fig:VIwithoutPPP}
\end{subfigure}
\medskip
\begin{subfigure}{0.3\textwidth}
  \centering
  \includegraphics[angle=-90,width=.8\linewidth]{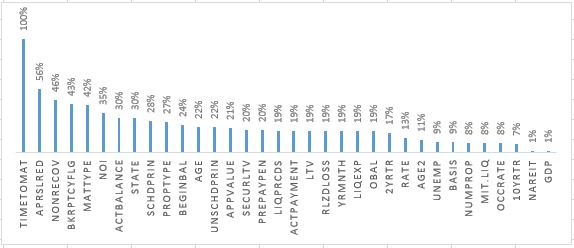}\quad
  \caption{VI without Balloon Payment}
  \label{fig:VIwithoutBal}
\end{subfigure}
\begin{subfigure}{0.3\textwidth}
  \centering
  \includegraphics[angle=-90,width=.9\linewidth]{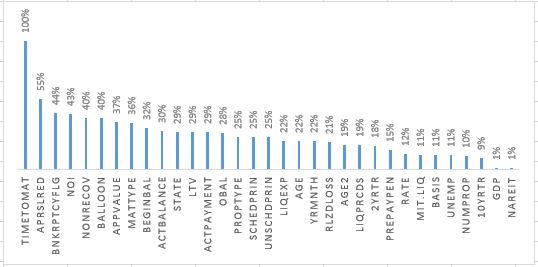}\quad
  \caption{VI without Occupancy}
  \label{fig:VIwithoutOcc}
\end{subfigure}
\begin{subfigure}{0.3\textwidth}
  \centering
  \includegraphics[angle=-90,width=.975\linewidth]{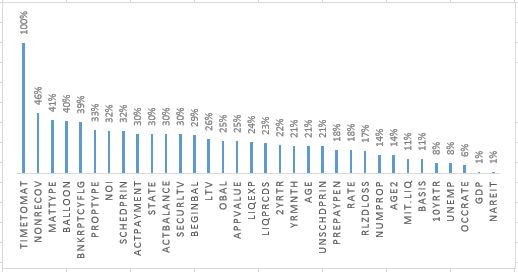}
  \caption{VI without Appraisal Reduc}
  \label{fig:VIwithoutAppRed}
\end{subfigure}
\caption{These are Variable Importance (VI) charts, leaving one out. I test the order of variable importance using Gedeon's method for DNN and exclude the key variables one at a time to test the strategic importance of NOI over LTV. NOI supercedes LTV in all of these VI charts, corroborating that NOI provides higher incentive than LTV during strategic default. Only the variables having more than 0.5\% marginal contribution are included in the variable importance visualizations for brevity.}
\label{fig:images1}
\end{figure}

\begin{figure}[!htbp]
\begin{subfigure}{0.3\textwidth}
  \centering
  \includegraphics[angle=-90,width=.8\linewidth]{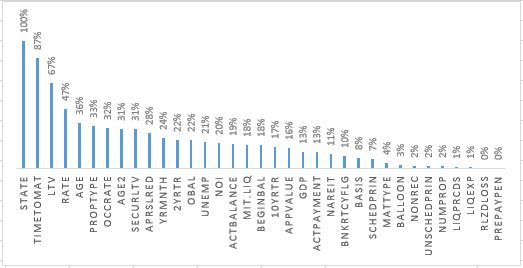}\quad
  \caption{VI in 2008: DRF}
  \label{fig:DRF2008}
\end{subfigure}%
\begin{subfigure}{0.3\textwidth}
  \centering
  \includegraphics[angle=-90,width=.8\linewidth]{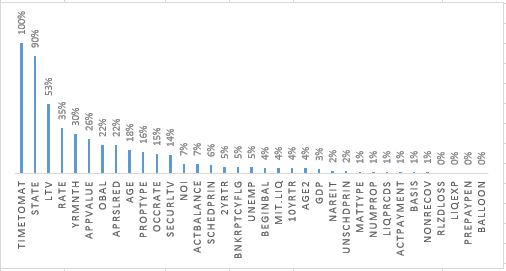}\quad
  \caption{VI in 2008: GBM}
  \label{fig:GBM2008}
\end{subfigure}
\begin{subfigure}{0.3\textwidth}\quad
  \centering
  \includegraphics[angle=-90,width=.8\linewidth]{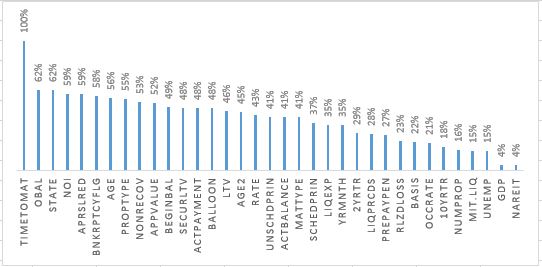}
  \caption{VI in 2008: DNN}
  \label{fig:DNN2008}
\end{subfigure}
\caption{Variable Importance is stress-tested during Financial Crisis across all non-parametric models. The ranking order of variables is maintained almost exactly for DNN even in the subsample leading upto the financial crisis. The other models for DRF and GBM are very sensitive to this catastrophic shock in 2008. This corroborates that the Variable Importance from Gedeon's method is robust and provides the order of incentives across variables while borrower chooses to default strategically. Only the variables having more than 0.5\% marginal contribution are included in the variable importance visualizations for brevity.}
\label{fig:Stress2008}
\end{figure}

Figure \ref{fig:images} reports variable importance across all six non-degenerate models (Lasso, Ridge, ordinal, DRF, GBM, DNN). NOI ranks above LTV only in the DNN. This is not an artifact of a particular specification: I verify the ranking's robustness in three ways. First, I re-estimate the DNN leaving out other strategic contractual variables one at a time -- year-month fixed effects, the prepayment penalty clause, the balloon payment indicator, occupancy, and appraisal reduction (Figure \ref{fig:images1}) -- and find that NOI continues to dominate LTV in every specification; removing year-month fixed effects in particular causes NOI's own importance to fall sharply, indicating that NOI's predictive power operates partly \emph{through} its correlation with macroeconomic timing rather than as a purely idiosyncratic loan-level signal. Second, I re-estimate all non-parametric models on data through June 2006 only and evaluate variable importance out-of-sample through the 2008 financial crisis (Figure \ref{fig:Stress2008}): the DNN's importance ranking is essentially unchanged, while the DRF continues to rank LTV well above NOI, and the GBM -- as a greedy, locally-optimizing algorithm -- is comparatively unstable, finding more local optima favoring LTV after boosting. Third, Section \ref{sec:covid} repeats this exercise for the COVID-19 shock to commercial real estate with materially the same result.

\begin{figure}[!htbp]
	\centering
	\begin{subfigure}{0.48\textwidth}
		\centering
		\includegraphics[angle=0,width=.8\linewidth]{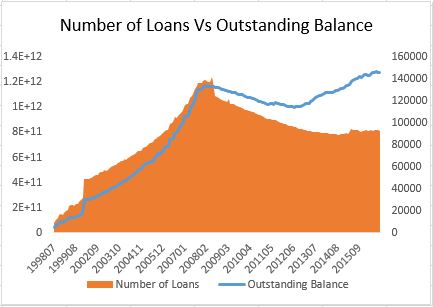}
		\caption{Number of Loans vs. Outstanding Loan Balance}
		\label{NumObalNew}
	\end{subfigure}
	\begin{subfigure}{0.48\textwidth}
		\centering
		\includegraphics[angle=0,width=.8\linewidth]{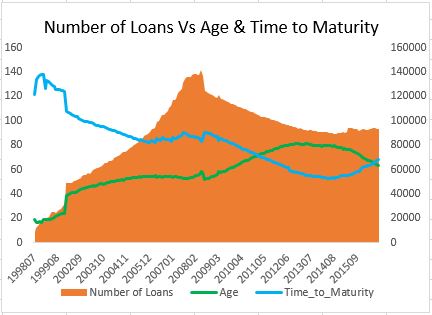}
		\caption{Number of Loans vs. Age and Time to Maturity}
		\label{NumTimeToMatAge}
	\end{subfigure} \\
	\begin{subfigure}{0.48\textwidth}
		\centering
		\includegraphics[angle=0,width=.8\linewidth]{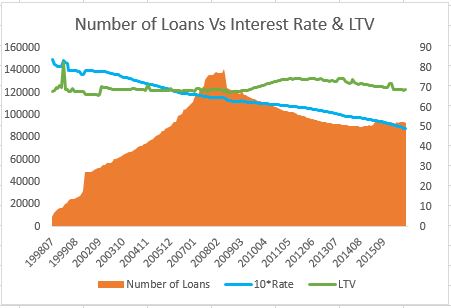}
		\caption{Number of Loans vs. Interest Rate and LTV}
		\label{NumRateLTV}
	\end{subfigure}
	\begin{subfigure}{0.48\textwidth}
	\centering 
	\includegraphics[angle=0,width=.8\linewidth]{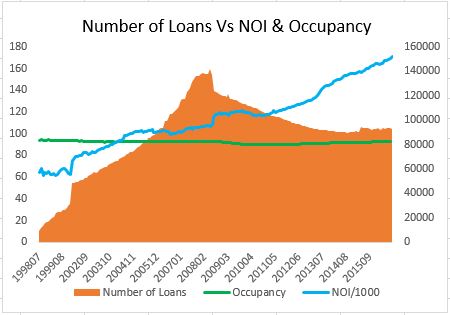}
	\caption{Number of Loans vs. NOI and Occupancy}
		\label{NumNOIOcc}
	\end{subfigure}
 	\caption{I provide evidence from the Trepp data in Figure \ref{NumObalNew} that from 2012, the number of loans have remained flat but the outstanding balance of loans have steadily increased until 2016. Figure \ref{NumTimeToMatAge} furthers the narrative. From mid-2014, the age of the loans is decreasing and the time-to-maturity is increasing. This could 
mean that from mid-2014, there are an equal number of originations to the number of maturing loans. But the fact that the Outstanding Balance is increasing in this entire period could only mean that the same loans are getting rolled over to new contracts, when balloon payments are missed during maturity. Figure \ref{NumRateLTV} clearly shows that LTV (widely used in previous literature and used by most banks/asset managers for credit risk calculations) is flat throughout the data horizon. The interest rate is decreasing almost monotonically in the data and there seems to be no sensitivity of LTV to interest rate. Figure \ref{NumNOIOcc} corroborates that the NOI monotonically increases in the data and the occupancy is almost 100\% in the entire data. So, there may be strategic saving of internal cash flow from income producing properties.}
		\label{fig: MotivatingDiagrams}
\end{figure}

Two descriptive patterns from the raw data (Figure \ref{fig: MotivatingDiagrams}) motivate why this ordering should not be surprising. From 2012 onward, the number of outstanding loans in the panel is roughly flat while the aggregate outstanding balance rises steadily through 2016 -- consistent with a stock of loans that are being rolled over into new contracts, rather than cured or resolved, when balloon payments come due. Over the same period, LTV is essentially flat despite a near-monotonic decline in interest rates, suggesting that commercial borrowers may be actively \emph{targeting} a stable LTV -- since banks and asset managers use LTV at origination, and contemporaneous LTV thereafter, as their primary underwriting and loss-given-default metric -- rather than passively responding to it. Over the same window, NOI rises monotonically and occupancy remains close to 100\%, consistent with strategic retention of internal cash flow from income-producing properties rather than mechanical liquidity constraints.

\subsection{The 2008 Financial Crisis and the 2020--2021 COVID-19 Shock}
\label{sec:covid}

To assess whether the NOI-over-LTV ranking is a sample-period artifact, I estimate the DNN on data through November 2019 -- prior to the first identified COVID-19 cases -- and evaluate variable importance and classification accuracy on a holdout running through September 2020. NOI continues to rank above LTV, and out-of-sample classification accuracy is comparable to the pre-pandemic benchmark (Appendix Table \ref{Training from Jan 2017-Nov2019, test Dec 2019 - Sept 2020}), whereas a model trained through February 2020 -- after the shock had begun to be priced in but before its full severity was apparent -- shows markedly worse out-of-sample accuracy, consistent with a genuine, unanticipated shock rather than a slow-moving trend the model could have extrapolated. Splitting the COVID-19 sample by property type (Appendix Figure \ref{Impact of COVID-19 across industries}) shows the hotel sector migrating rapidly from current to 30--60 and then 60--90 days delinquent starting in May 2020, with a marked rise in loans beyond 120 days delinquent (effectively becoming ``limbo'' loans in the sense of \citet{LimboLinda2015}) by August 2020; the office sector instead shows a persistent rise in non-performing, maturity-proximate loans, consistent with borrowers attempting to roll over balloon payments amid historically low interest rates rather than resolving delinquency outright.

Notably, NOI's importance advantage over LTV disappears specifically for loans that have already resolved to real-estate-owned (REO) or foreclosure status (Appendix Section \ref{Determinants of Each Delinquency Class}), for the mechanical reason that NOI is no longer a meaningful state variable once the lender has taken back the property or foreclosure proceedings are underway. This placebo-like pattern is reassuring: NOI's importance is concentrated exactly where the theoretical model in Section \ref{sec:Theory} predicts it should matter -- while the borrower still controls the property and the decision to keep paying remains live -- and vanishes once that decision has already been resolved.

These results are consistent with, and help explain the mechanism behind, two recent findings on COVID-era CRE distress. \citet{GriffinPriest2023} show that underwritten NOI on post-2010 (``CMBS 2.0'') loans is systematically overstated relative to realized property income, and that the degree of overstatement is highly predictive of pre- and COVID-period loan distress -- consistent with this paper's finding that reported NOI, not just LTV, carries first-order information about which loans will ultimately default, and suggesting that some of the ``strategic'' NOI-conditional default this paper documents may partly reflect borrowers or originators whose reported income was optimistic to begin with. \citet{GuptaMittalVanNieuwerburgh2026} independently document a severe and persistent post-pandemic decline in office-sector lease revenue and occupancy, driven by the shift to remote work; the pronounced office-sector delinquency migration this paper documents in Figure \ref{Impact of COVID-19 across industries} is a loan-performance counterpart to the valuation decline they estimate directly from lease-level data.

\section{Conclusion}
\label{sec:conclusion}

This paper uses a deep neural network on a large, proprietary panel of commercial mortgages to separate two competing, observationally similar explanations for delinquency -- liquidity-constrained default and strategic default -- that standard parametric models are not well equipped to distinguish. Conditional on delinquency, property income (NOI) is a more important and more robust determinant of the default rate than leverage (LTV), a ranking that survives leave-one-variable-out robustness checks, an out-of-sample stress test through the 2008 financial crisis, and an analogous test through the 2020--2021 COVID-19 shock to commercial real estate. A bivariate analysis of default rates across LTV and NOI buckets recovers economically interpretable thresholds -- roughly LTV of 0.20, 0.60, and 0.82, and NOI near the fourth and sixth percentiles of the Trepp distribution -- that separate liquidity-constrained from strategic default patterns, and that a placebo-style check (NOI's importance vanishing once a loan reaches REO or foreclosure status) helps validate as economically, rather than purely statistically, meaningful.

These findings matter for at least three audiences. For CMBS investors and rating agencies, they suggest that underwriting and surveillance frameworks built primarily around LTV may materially understate default risk for loans with erratic or declining property income, even when leverage appears moderate, and may simultaneously overstate risk for highly levered but cash-flow-strong properties outside the strategic-default range identified here. For special servicers, the LTV--NOI threshold structure offers a concrete, ex ante screening tool for distinguishing loans that are likely to cure with forbearance (liquidity-constrained) from loans where modification is unlikely to prevent eventual loss (strategic), which is directly relevant to renegotiation and loss-mitigation strategy. For policymakers and regulators, the results caution against countercyclical capital or forbearance policies calibrated solely to leverage metrics, since a meaningful share of the default risk this paper documents operates through the income channel instead.

Several limitations point toward future work. The DNN's variable-importance rankings, while stable across sub-periods and robust to leave-one-out checks, remain a reduced-form summary of a fitted model rather than a causal estimate in the sense of an instrumented or quasi-experimental design; a natural extension would combine this paper's threshold-identification approach with an explicit, loan-level instrument for NOI shocks where one can be found (for example, exploiting idiosyncratic lease-expiration timing). The theoretical model in Section \ref{sec:Theory} is deliberately spare and does not solve explicitly for the borrower's optimal stopping rule; a fully specified optimal-stopping model calibrated to the estimated thresholds would sharpen the paper's structural interpretation and could, in future work, be combined with the frailty-model machinery of \citet{Duffie:2009} and \citet{Yildirim2008} to separate long-run survivorship from default timing within the strategic-versus-liquidity-constrained framework developed here. Finally, extending the sample beyond CMBS-securitized loans to balance-sheet CRE lending -- where NOI reporting is typically unavailable -- remains an open data challenge, and one that likely requires new data-collection efforts rather than purely methodological ones.

\pagebreak
\bibliographystyle{plainnat}
\bibliography{ref}

\newpage
\begin{appendices}
\setcounter{table}{0}
\setcounter{section}{0}
\setcounter{figure}{0}
\renewcommand{\thetable}{A\arabic{table}}
\renewcommand{\thesection}{A\arabic{section}}
\renewcommand{\thefigure}{A\arabic{figure}}
\renewcommand{\thesection}{A}

\tableofcontents
\addtocontents{toc}{\protect\setcounter{tocdepth}{3}}  

\clearpage

\subsection{Loan to Value and Net Operating Income}
\label{Loan to Value and Net Operating Income}

I take a deeper dive and investigate LTV in the following way:
\begin{equation}
LTV_t = \frac{AOB_{t-1} + DS_t + B_T}{MV_t - AR_t}
\end{equation}
where $AOB_t$ is the Outstanding Balance at time t-1 that is amortized , $DS_t$ is the scheduled payment due for servicing the debt obligation at time t, $B_T$ is the Balloon Payment due at maturity, $MV_t$ is the Market value of the property/properties at time t (which varies significantly with respect to macroeconomic conditions and spatial/location 
context) for which the mortgage has been issued, $AR_t$ is the Appraisal Reduction at time t. 

AOB remains consistent, since, prepayment penalty clauses discourage voluntary curtailment/full prepayment. SP obligations are not met both when the borrower is cash-constrained and also when the borrower chooses to strategically default. 
Proximity to balloon payment at maturity further complicates the endogenous behavior of the commercial borrowers towards maturity of the loan. The market value of a property is a function of the macro-economic factors like state GDP, Unemployment Rate, geographical location, 2 Year and 10 Year Treasury Rates. Until valuation is obtained, Appraisal Reduction Amount (ARA) may be calculated based on the scheduled principal balance or some other formula as defined in the servicing agreement.

NOI calculation involves the following key variables. Potential Rental Income assumes zero vacancy or could be based on a rental market analysis. Vacancy losses are realized when tenants vacate the property and/or tenants default on their lease obligations. Total Operating Expenses on an Investment Property could include "Property Taxes, Rental Property Insurance, Property Management Fees, Maintenance and Repairs, Miscellaneous Expenses, etc. Debt service, depreciation, leasing commissions, tenant improvements, repairs to wear and tear, income taxes, and mortgage interest expenses are not included in the calculation of net operating income". This is because NOI is property-specific devoid of other investor or borrower-specific expenses. NOI helps calculate Cap Rate (property's potential rate of return), ROI, Debt Coverage Ratio, Cash Return on Investment. NOI provides an estimate a property's ongoing operating revenue. NOI analysis can be manipulated from the choice to accelerate or defer certain expenses. The NOI of a property can change depending on the property management. Because other expenses are not considered in NOI, the real cash flow from a property may differ net other expenses. Further projected rents cannot be used to calculate NOI when rents differ from market rents.

\subsection{Multinomial Logistic Model}
\label{Appendix: Multinomial Logistic Model}

In a Multinomial Logistic Model, log-odds of each delinquency state with respect 
to the "Current" state assumes a linear specification. The odds that a loan has 
a delinquency classes j as opposed to the baseline, depending only on individual 
loan-specific covariates is defined as:
\begin{equation}
\frac{Pr(Y_i=j|Z_i=z)}{Pr(Y_i=0|Z_i=z)} = exp(Z^{'}\gamma_j)
\end{equation}
the choice $Y_i$ takes on non-negative, un-ordered integer values $Y_i \in \{0, 1, . . . , J\}$. Multinomial logistic regression does not assume normality, linearity, or homoskedasticity; it has a well-behaved likelihood function, a special case of conditional logit. A more powerful alternative to multinomial logistic regression is discriminant function analysis which requires these assumptions are met. Multinomial logistic regression also assumes non-perfect separation.

The Independence of Irrelevant Alternatives (IIA) assumption inherent in Multinomial Logistic Model implies that adding or deleting alternative outcome categories does not affect the odds among the remaining outcomes. 
\begin{equation}
Pr(Y_i=j|Y_i\in\{j,l\}) = \frac{Pr(Y_i=j)}{Pr(Y_i=j)+Pr(Y_i=l)} = \frac{exp(X_{ij}^{'}\gamma) }{exp(X_{ij}^{'}\gamma) + exp(X_{il}^{'}\gamma)}
\end{equation}
This can be tested by the Hausman-McFadden test. There are alternative modeling methods, such as alternative-specific multinomial probit model, or nested logit model to relax the IIA assumption. 

\subsubsection{Independence of Irrelevant Alternatives}

\textbf{Multinomial Logistic Model} assumes Independence of Irrelevant Alternatives (IIA). The following Finite State Automaton details all possible transitions so that the above arguments can be visualized.

\begin{tikzpicture}[->,>=stealth',shorten >=1pt,auto,node distance=3cm,
        scale = 0.6,transform shape]

  \node[state] (Current) {$Current$};
  \node[state] (30DPD) [right of=Current] {$30DPD$};
  \node[state] (60DPD) [right of=30DPD] {$60DPD$};
  \node[state] (90DPD) [right of=60DPD] {$90DPD$};
  \node[state,accepting] (120DPD) [right of=90DPD] {$120DPD$};

\path (Current) edge   [bend left=20]           node {$ $} (30DPD)
(Current) edge     [loop above]         node {$ $} (Current)
(30DPD) edge     [bend left=20,thick]         node {$ $} (Current)
(30DPD) edge     [loop above]         node {$ $} (30DPD)
(30DPD) edge       [bend left=20]       node {$ $} (60DPD)
(60DPD) edge        [bend left=20]      node {$ $} (90DPD)
(60DPD) edge      [bend left=20,thick]        node {$ $} (30DPD)
(60DPD) edge          [bend left=40,thick]    node {$ $} (Current)
(60DPD) edge         [loop above]     node {$ $} (60DPD)
(90DPD) edge         [bend left=20]     node {$ $} (120DPD)
(90DPD) edge         [loop above]     node {$ $} (90DPD)
(90DPD) edge       [bend left=20,thick]       node {$ $} (60DPD)
(90DPD) edge        [bend left=40,thick]      node {$ $} (30DPD)
(90DPD) edge             [bend left=60,thick] node {$ $} (Current)
(120DPD) edge            [loop above]  node {$ $} (120DPD)
(120DPD) edge           [bend left=20,thick]   node {$ $} (90DPD)
(120DPD) edge             [bend left=40,thick] node {$ $} (60DPD)
(120DPD) edge             [bend left=60,thick] node {$ $} (30DPD)
(120DPD) edge              [bend left=80,thick] node {$ $} (Current);

\end{tikzpicture}

Clearly, the borrower would like to stick with the first choice, as the second choice classifies him/her in the default category and is detrimental for her creditworthiness from a lender's perspective. Now suppose, one more choice for being in \textbf{30 days to 60 days of delinquency} is given to the borrower, s/he may choose to rather be in this new state instead of \text{less than 30 days of delinquency} and may \textbf{strategically} miss one payment if there is a great investment opportunity for him/her in that one month horizon. In fact, none of the models (except Ordinal Bayes) can distinguish these three classes (\textbf{W0\_30D}, \textbf{W30\_60D} \& \textbf{W60\_90D}) and considers all of them as \textbf{Current Loans} in Table \ref{tab:Test}.

The borrower can undertake this decision as she/she is already some days in delinquency and she/she wouldn't mind going to the next bucket until she/she falls in the bucket for \textbf{90 days to 120 days of delinquency}. In this situation, the borrower's creditworthiness doesn't change that much from a lender's perspective. hence, the odds for being in the \textbf{"less than 30 days delinquency"} to being in the classes of \textbf{90 days to 120 days of delinquency} will change drastically in the presence of this new choice of being in \textbf{30 days to 60 days of delinquency}. hence the IIA assumption is clearly violated. 

Also the marginal effect of features towards classifying the response set does not have a clear interpretation in terms of sensitivity and directionality. I list the co-efficients of Multinomial Logistic Model for the sake of completeness. (Table \ref{Co-efficients of Multinomial Logistic Regression})

\begin{table}[!htbp]
\tiny
\centering
\caption{\textbf{Coefficients of Multinomial Logistic Regression:} The marginal effect of features towards classifying the response set does not have a clear interpretation in terms of sensitivity and directionality. I list the co-efficients of Multinomial Logistic Model for the sake of completeness and only list the standard errors (SE) of the 3 statistically significant variables (Rate, LTV, Unemployment) that I use to demonstrate the counterintuitive interpretation in terms of default. (Table \ref{Co-efficients of Multinomial Logistic Regression}) }
\label{Co-efficients of Multinomial Logistic Regression}
\begin{tabular}{rlrrrrrrrrrrrrrr}
\hline
& names & W0\_30D & W30\_60D & W60\_90D & W90\_120D & PrfMatBal & NPfMtBl & B120D \\ 
\hline
1 & Intercept & -24.83 & -230.36 & -223.45 & -319.34 & 591.88 & -170.20 & 272.02\\ 
2 & State & -0 & 0.69 & 0.30 & 0.99 & -2.12 & -0.76 & -0.08 \\ 
52 & PropType & -0.39 & 0.65 & -0.12 & 0.61 & 0.55 & 0.62 & 0.49 \\ 
60 & MatType & 0.23 & -4.82 & -2.87 & -3.79 & 7.32 & -57.90 & -6.62 \\ 
61 & Balloon & 0.33 & -5.39 & -3.48 & -4.05 & 5.79 & -53.02 & -5.83 \\ 
62 & NonRecov & -1.59 & -0.94 & -0.75 & 0.69 & -3.13 & 0.53 & 0.43 \\ 
63 & BnkrptcyFlg & -0.15 & 0.02 & 0.19 & 1.10 & -0.78 & -1.88 & -0.09 \\ 
64 & BeginBal & -3.39 & -2.13 & -3.27 & 1.56 & -0.67 & -8.03 & 5.99 \\ 
65 & Obal & -1.18 & 21.62 & 19.06 & -41.82 & 18.39 & -7.16 & -115.35 \\ 
66 & \textbf{Rate} & $\textbf{-17.31}^{***}$ & 29.77 & 29.74 & 34 & $\textbf{-27.34}^{***}$ & 8.05 & 32.65 \\ 
    & \textbf{SE} & \textbf{0.21} & 0.35 & 0.35 & 0.40 & \textbf{0.33} & 0.10 & 0.39 \\ 
67 & SchedPrin & 134.71 & 153.33 & 54.63 & 89.44 & 156.85 & 173.07 & 41.50 \\ 
68 & Payment & 44.44 & -361.30 & -6.64 & -37.43 & 24.03 & -1.95 & 57.25 \\ 
69 & UnschdPrin & 45.46 & -133.66 & 70.57 & -47.70 & 57.08 & -71.93 & -172.29 \\ 
70 & PrePayPen & 11.65 & -253.05 & -103.59 & -61.27 & -44.45 & -370.52 & 16.91 \\ 
71 & ActBalance & 2.25 & 0.94 & 2.73 & -2.03 & -0.28 & 8.07 & -4.28 \\ 
72 & Liqprcds & -75.90 & 76.39 & 35.06 & 131.46 & 8.39 & 179.17 & 165.38 \\ 
73 & Liqexp & 448.71 & -1139.31 & -2653.88 & -518.47 & -124.37 & -1035.34 & 72.08 \\ 
74 & RlzdLoss & 69.41 & 26.56 & -48.44 & -42.75 & 39.97 & -43.45 & -58.66 \\ 
75 & NumProp & -2.03 & 3.72 & 4.96 & 16.78 & 5.49 & 4.55 & -23.82 \\ 
76 & AprslRed & -101.36 & 8.64 & 11.74 & 66.68 & -20.31 & 49.31 & 87.47 \\ 
77 & SecurLTV & 5.65 & 5.77 & 5.39 & 3.49 & 3.20 & 3.68 & -4.19 \\ 
78 & NOI & 159.73 & -7.38 & -39.41 & 27.03 & 124.60 & -77.96 & -149.62 \\ 
79 & \textbf{LTV} & $\textbf{-6.55}^{***}$ & -3.10 & -2.39 & 0.03 & -3.56 & -2.21 & 0.77 \\ 
     & \textbf{SE} & \textbf{1.11} & 0.53 & 0.41 & 0.01 & 0.61 & 0.38 & 0.13 \\ 
80 & AppValue & -2.27 & 3.02 & 4.01 & 7.02 & -201.11 & 11.14 & 9.66 \\ 
81 & OccRate & 0.42 & -3.10 & -3.55 & -3.51 & -1.27 & -2.28 & -2.84 \\ 
82 & Basis & -7.14 & -2.13 & -1.89 & 0.28 & 19.61 & 48.04 & 5.77 \\ 
84 & Age & -6.55 & -5.95 & -4.29 & -4.77 & -2.70 & 6.02 & 8.03 \\ 
85 & Age2 & 16.16 & 0.62 & -2.87 & -1.52 & 12.32 & -10.49 & -8.52 \\ 
86 & \textbf{Unemp} & $\textbf{-1.93}^{***}$  & 17.51 & 21.77 & 18.87 & 25.90 & 21.55 & $\textbf{-12.23}^{***}$ \\ 
    & \textbf{SE} & \textbf{0.05} & 0.43 & 0.53 & 0.46 & 0.64 & 0.53 & \textbf{0.30} \\ 
87 & GDP & -0.84 & 0.75 & 0.27 & 3.30 & 8.21 & 2.35 & -6.66 \\ 
88 & 2YrTr & 4.89 & -7.05 & -9.52 & -34.94 & -5.31 & -50.57 & -3.23 \\ 
89 & 10YrTr & 10.94 & 18.14 & 18.82 & 37.29 & -20.60 & 7.96 & 15.26 \\ 
90 & NAREIT & 0.10 & 0.13 & 0.32 & 0.20 & 0.02 & 0.41 & 0.17 \\ 
91 & MIT.Liq & 0.29 & -0.12 & -0.35 & -0.31 & 0.27 & -0.34 & 3.71 \\ 
92 & TimeToMat & 4.69 & 4.63 & 4.43 & 4.57 & -101.20 & -140.81 & 2.91 \\ 
\hline
\end{tabular}
\end{table}

\subsection{Distributed Random Forest}
\label{Appendix: Distributed Random Forest}

Recursive partitioning, a critical data mining tool, helps in exploring the structure of a data set. This section provides a brief overview of CART modeling, conditional inference trees, and random forests.


Random Forests are developed by aggregating decision trees and can be used for both classification and regression. Each tree is a weak learner created from bootstrapping from subset of rows and columns. More trees will reduce the variance.It alleviates the issue of overfitting, can handle a large number of features. It helps with feature selection based on importance. It is user-friendly with two parameters: number of trees (default 500) and variables randomly selected as candidates at each split, $\sqrt{ntree}$ for classification and
$ntree/3$ for regression. "Out Of Bag Error" is estimated for each bootstrap iteration and related tree.



R's randomForest splits based on the Gini criterion and H2O trees are split based on reduction in Squared Error (even for classification). H2O also uses histograms for splitting and can handle splitting on categorical variables without dummy (or one-hot) encoding. Also, R's randomForest builds really deep trees, resulting in pure leaf nodes, leading to constant increments in prediction and ties and hence relatively lower AUC.The trees in H2O's random forest aren't quite as deep and therefore aren't as pure, allowing for predictions that have some more granularity to them and that can be better sorted for a better AUC score.

CART models an outcome $y_i$ for an instance i as:

\begin{equation}
    y_i = f(x_i) = \sum_{m=1}^{M} c_m I{x_i \in R_m}
\end{equation}

where each observation $x_i$ belongs to exactly one subset $R_m$, $c_m$ is the mean of all training observations in $R_m$.


\subsection{Gradient Boosting Machine}
\label{Appendix: Gradient Boosting Machine}

GBM \cite{Greedyfunction} creates an ensemble \cite{Kuncheva2003} of weak prediction models in stages and utilizes a differentiable loss function. Boosting trees does increase accuracy, but at the cost of speed and meaningful interpretability. 

At each step m, $1 \leq m \leq M$ of gradient boosting, an estimator $h_m$ is computed from the residuals of the previous model predictions. Friedman (2001) proposed regularization by shrinkage:
\begin{equation}
    F_{m} (x) = F_{m-1} (x) + \nu \gamma_{m} h_{m} (x)
\end{equation}
where $h_{m} (x)$ represents a weak learner of fixed depth, $\gamma_m$ is the step length and $\nu$ is the learning rate or the shrinkage factor. XGBoost \citet{DBLP:journals/corr/ChenG16} is a faster and more accurate implementation of the Gradient Boosting algorithm \citet{DBLP:journals/corr/abs-1801-07384}.

\subsection{Deep Neural Network}

\begin{figure}[!htbp]
	\centering
	\includegraphics[width=6.8in]{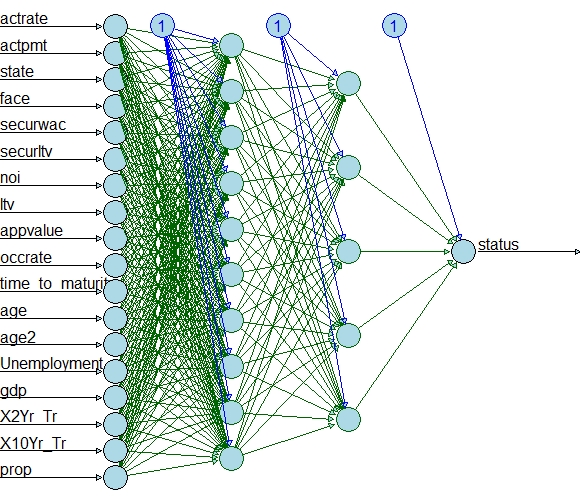}
	\caption{A few variables out of 33 are droppped in this visualization due to several missing values. However, they have been included in the actual implementation of the Feedforward DNN from h2o package. For visualization, the package neuralnet throws an error for variables with several missing values. But this diagram clearly shows how DNN works with hidden layers and how variables have weights during linear combination in each layer and then fed forward using non-linear activation functions and finally to the outcome variable during classification.}
	\label{fig:Deep Neural Network}
\end{figure}
\label{sec:Glossary}
\subsubsection{Deep Neural Network for CMBS}
\label{sec:deep}
The purpose for a Deep Neural Network (DNN) is bourne out of the need to have transparency and accountability \citet{Albanesi}. By the very nature of the DNN, I do not have to add interaction terms in the specification of the model, especially in the case of high dimensional data. The sequential layers embody highly non-linear and non-trivial interaction among the variables and capture several latent fundamental features in the process. The causal interpretation of the covariates both in default \citet{Kvamme2018PredictingMD} and prepayment calculations have not been explored in details. The broader impact could be traced out by improved allocation of credit and aid in policy design (macroprudential, bankruptcy, foreclosure, etc.).

With the provision of enough hidden units, a neural network can mimic continuous functions on closed and bounded sets really well \citet{Hornik1991}, vis-a-vis the product and division of relevant features and their
interactions. More layers, and not more units in each layer, learns atures of greater complexity. Deep neural networks \ref{fig:Deep Neural Network}, with three or more hidden layers, require exponentially fewer units than shallow networks or logistic regressions with basis functions; see  \cite{MontuferPAscanuChoBengio2014} and \cite{BengioCourvilleGoodfellow2016}.

Deep Neural Networks (DNN) have an extensive set of current applications like: System identification and control (e.g., vehicle control, trajectory prediction, etc.), 
Game-playing and decision making (e.g., chess, poker, etc.), Pattern and sequence 
recognition (e.g., radar systems, face identification, signal classification, speech/image recognition, etc.), 
Medial diagnosis and finance (e.g., automated trading systems, cancer diagnosis, etc.).

\subsection{Hyperparameter Tuning and Grid Search}
\label{sec:Hyperparameter}

Hyper-parameter tuning with Random Grid Search (RGS) tests different combinations of hyper-parameters to find the optimal choice based on accuracy, without overfitting. 

Hyperparameters can be divided into 2 categories:

\begin{itemize}
	\item \textbf{Optimizer hyperparameters}
	\item \textbf{Model Specific hyperparameters}
\end{itemize}

My model hyperparameters are: \emph{score\_training\_samples} = 6125796, \emph{epoch} = 60000, \emph{hidden} = c(30,20,10,7), \emph{hidden\_dropout\_ratios} = c(0.01, 0.01, 0.01,0.01), \emph{momentum\_start} = 0.5, \emph{momentum\_ramp} = 100, \emph{momentum\_stable} = 0.99, \emph{missing\_values\_handling} = "Skip", \emph{initial\_weight\_distribution} ="Uniform", \emph{nesterov\_accelerated\_gradient} = TRUE, \emph{activation} = "RectifierWithDropout", \emph{nfolds} = 10, \emph{fold\_assignment} = "Stratified", \emph{keep\_cross\_validation\_predictions} = FALSE, \emph{variable\_importances} = TRUE, \emph{adaptive\_rate} = FALSE, \emph{l1} =1e-5, \emph{l2} =1e-5, \emph{export\_weights\_and\_biases} = FALSE, \emph{mini\_batch\_size} = 128, \emph{loss} = "CrossEntropy", \emph{distribution} = "AUTO", \emph{balance\_classes} = T, \emph{max\_after\_balance\_size} = 1, \emph{rate} = 05, \emph{rate\_annealing} = 1e-06, \emph{rate\_decay} = 1, \emph{stopping\_metric} = "MSE", \emph{seed} = 1122.

\subsection{Class Imbalance Problem}
\label{sec:SMOTE}

Most classifiers are unable to distinguish minor classes \cite{Kuncheva2003} and are heavily influenced by major classes, e.g., the conditional probability of minor classes are underestimated  in a logistic regression \cite{KingZeng2001}, Tree based classifiers, and KNN yield high recall but low sensitivity when the data set is extremely unbalanced \cite{DBLP:conf/pkdd/2008-1}. There are a plethora of techniques to balance the data, e.g., oversampling, under-sampling and Synthetic Minority Oversampling Technique (SMOTE)  proposed by \cite{DBLP:journals/jair/ChawlaBHK02}. 

\section{COVID-19 Results without internalizing 2008 Crisis}
\label{COVID-19 Results}
After cleaning the data, I have 1,315,421 observations from Jan 2017 - Sep 2020. I first try training the DNN model with data from Jan 2017 - Feb 2020 and confirm that NOI is more important than LTV leading upto the COVID-19 pandemic in Table \ref{Training from Jan 2017-Feb2020, test March2020 - Sept 2020}. However, consistent with my conjecture that Commercial Mortgage delinquency leads a financial (in this case, induced by a global health crisis), the out-of sample predictions are inaccurate. Then I train the DNN model from Jan 2017 - Nov 2019, as the first cases of COVID-19 were identified in Wuhan, China during December, 2019. I still find that NOI is more important than LTV in Table \ref{Training from Jan 2017-Nov2019, test Dec 2019 - Sept 2020} and also provide evidence about other variables which become important in co-determining the default behavior. I find similar accuracy in predictions in Table \ref{Confusion Matrix: Training from Jan 2017-Nov2019, test Dec 2019 - Sept 2020} compared to Table \ref{tab:Test}. This solidifies my previous and I can indeed claim that DNN extracts the inherent structural relationship among the covariates and can robustly predict even during several financial crises. 

\subsection{Results by Industry during COVID-19}
\label{Results by Industry during COVID-19}

\begin{figure}[!htbp]
\centering
\includegraphics[angle=90, width=8in]{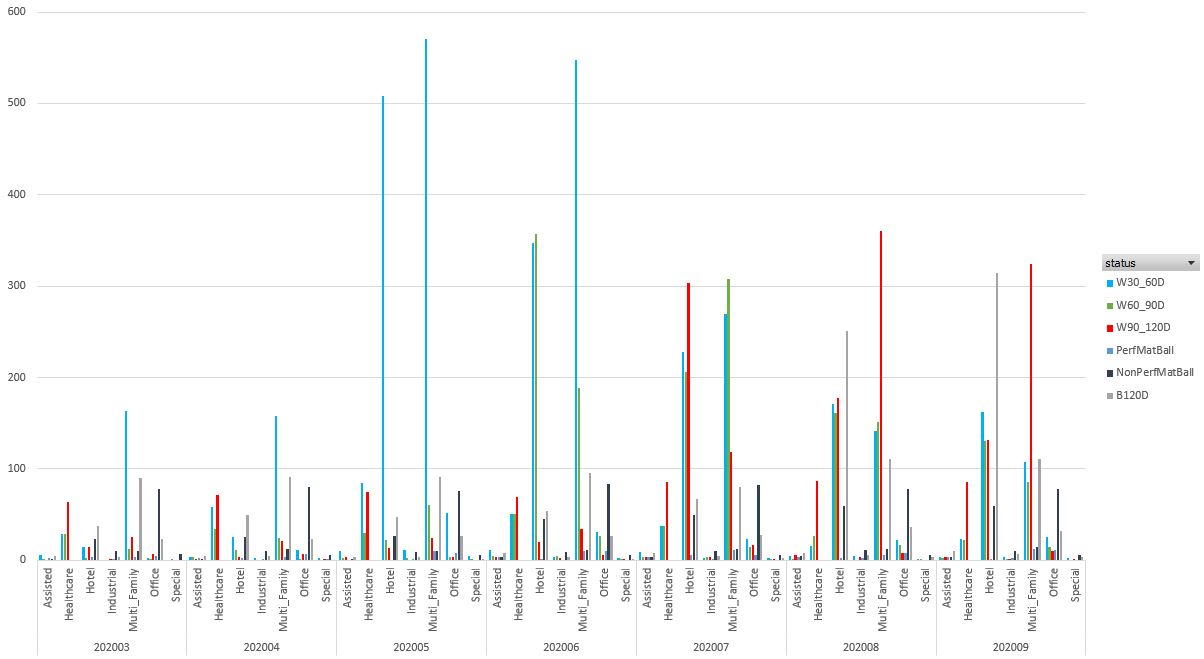}
\caption{Impact of COVID-19 across industries}
\label{Impact of COVID-19 across industries}
\end{figure}

I track the evolution of the number of loans in different delinquency classes across time (Dec 2019 - Sep 2020) and across industries (Assisted, Healthcare, Hotel, Industrial, Multi-family, Office, Special) in Figure \ref{Impact of COVID-19 across industries}. I clearly see the hotel industry being massively displaced. I see enormous number of commercial mortgage loans in the hotel industry degrade from \emph{W0\_30D} to \emph{W30\_60D} (orange bar) from May, 2020. There is some degradation from \emph{W30\_60D} (orange bar) to \emph{W60\_90D} (ash-color bar). These loans become limbo from August 2020 as evident from the rise in dark-blue bar of \emph{B120D} loans. The office space has also seen massive cashflow shortages from the lack of business activities and inability of payments therefrom for tenants in big cities. This is evidenced by constant green bar (Non-Performing Mature Balloon Loans) which are trying to roll-over the loan contract and some are successful from the lax underwriting standards due to historical levels of low interest rate monetary policy.

\subsection*{Determinants of Each Delinquency Class}
\label{Determinants of Each Delinquency Class}

I zoom in to individual delinquency classes and find the determinants of each adverse delinquency class using DNN. In Table \ref{BK: Variable Importance}, I still find NOI higher than LTV for DNN trained on Jan 2017 - Nov 2019 and misclassification error is almost 0\%. For Real-Estate-owned (REO) loans, the misclassification error is again 0\% in Table \ref{REO: Confusion Matrix}, but NOI is no longer higher than LTV in Table \ref{REO: Variable Importance} as there is no NOI when the lender takes back the property. For foreclosure, again the misclassification error is again 0\% in Table \ref{Foreclosure: Confusion Matrix} and similarly NOI is no longer higher than LTV in Table \ref{Foreclosure: Variable Importance} as foreclosure proceedings are lengthy processes which start after Bankruptcy Chapter 13 and there is no renegotiation to be done and either the loan gets resolved in court or leads to the terminal state of Bankruptcy Chapter 7 (liquidation). As shown in Subsection \ref{Results by Industry during COVID-19}, the behavior of W90\_120D and B120D loans are determined by industry heterogeneity. There is slightly higher error (in Tables \ref{W90120D: Confusion Matrix} and \ref{B120D: Confusion Matrix}) for these two uncertain delinquency states as the COVID-19 is an ongoing pandemic and still unfolding. 

\begin{table}[!htbp]
\tiny
\centering
\caption{I report the Variable Importance table for DNN (Deep Neural Network) across Delinquency Classes W0\_30D (Within 30 days), W30\_60D (Within 30-60 days), W60\_90D (Within 60-90 days), W90\_120D (Within 90-120 days), PrfMatBal (Performing Mature Balloon), NPrfMatBal (Non-Performing Mature Balloon), B120D (Beyond 120 days) with sample ranging from Jan 2017 - Feb 2020. NOI is higher than LTV, but accuracy reduces as the effect of COVID-19 was priced in by Feb 2020.}
\label{Training from Jan 2017-Feb2020, test March2020 - Sept 2020}

\begin{tabular}{rlrrr}
  \hline
 variable & relative\_importance  \\ 
  \hline
 state & 1 \\ 
 appvalue & 0.99 \\ 
 nonrecover & 0.94 \\ 
 bankruptcyflag & 0.90 \\ 
 securltv & 0.90 \\ 
 proptype & 0.89  \\ 
 2Yr\_Tr & 0.87 \\ 
 Unemployment & 0.83  \\ 
 pmtbas & 0.83  \\ 
 hasballoon & 0.82  \\ 
 maturitytype & 0.82 \\ 
 age & 0.80  \\ 
 face & 0.79 \\ 
 year\_month & 0.78 \\ 
 gdp & 0.76  \\ 
 10Yr\_Tr & 0.73  \\ 
 actrate & 0.73 \\ 
 \textbf{noi} & 0.72  \\ 
 age2 & 0.70 \\ 
 occrate & 0.68  \\ 
 securwac & 0.68  \\ 
 actpmt & 0.68 \\ 
 \textbf{ltv} & 0.67  \\ 
 time\_to\_maturity & 0.65  \\ 

   \hline
\end{tabular}

\end{table}

\begin{table}[!htbp]
\tiny
\centering
\caption{I report the Variable Importance table for DNN (Deep Neural Network) across Delinquency Classes W0\_30D (Within 30 days), W30\_60D (Within 30-60 days), W60\_90D (Within 60-90 days), W90\_120D (Within 90-120 days), PrfMatBal (Performing Mature Balloon), NPrfMatBal (Non-Performing Mature Balloon), B120D (Beyond 120 days) with sample ranging from Jan 2017 - Nov 2019. NOI is higher than LTV and accuracy increases as the effect of COVID-19 was not priced in by Nov 2019.}
\label{Training from Jan 2017-Nov2019, test Dec 2019 - Sept 2020}
\begin{tabular}{rlrrr}
  \hline
 variable & relative\_importance  \\ 
  \hline
 age2 & 1  \\ 
 age & 0.93  \\ 
 state & 0.90 \\ 
 time\_to\_maturity & 0.86  \\ 
 proptype & 0.78  \\ 
 face & 0.68  \\ 
 maturitytype & 0.64  \\ 
 nonrecover & 0.60  \\ 
 hasballoon & 0.60  \\ 
 actrate & 0.60  \\ 
 securwac & 0.59  \\ 
 actpmt & 0.57  \\ 
 bankruptcyflag & 0.56  \\ 
 pmtbas & 0.56 \\ 
 appvalue & 0.56  \\ 
 securltv & 0.55  \\ 
 year\_month & 0.52 \\ 
 \textbf{noi} & 0.51 \\ 
 2Yr\_Tr & 0.50  \\ 
 occrate & 0.45 \\ 
 \textbf{ltv} & 0.41  \\ 
 10Yr\_Tr & 0.39  \\ 
 Unemployment & 0.27  \\ 
 gdp & 0.07  \\ 
   \hline
\end{tabular}
\end{table}

\begin{table}[!htbp]
\tiny
\centering
\caption{I report the out of sample Confusion Matrix for DNN (Deep Neural Network) across Delinquency Classes W0\_30D (Within 30 days), W30\_60D (Within 30-60 days), W60\_90D (Within 60-90 days), W90\_120D (Within 90-120 days), PrfMatBal (Performing Mature Balloon), NPrfMatBal (Non-Performing Mature Balloon), B120D (Beyond 120 days) with training sample ranging from Jan 2017 - Nov 2019 and test sample from Dec 2019 - Sep 2020. Accuracy becomes comparable to Table \ref{tab:Test}.}
\label{Confusion Matrix: Training from Jan 2017-Nov2019, test Dec 2019 - Sept 2020}
\begin{tabular}{rrrrrrrrrl}
  \hline
 & \textbf{W0\_30D} & \textbf{W30\_60D} & \textbf{W60\_90D} & \textbf{W90\_120D} & \textbf{PrfMatBal} & \textbf{NPrfMatBal} & \textbf{B120D} & \textbf{Error} \\ 
  \hline
\textbf{W0\_30D} & 565681 & 10 & 81 & 370 & 0 & 3588 & 1290 & 1\% \\ 
  \textbf{W30\_60D} & 3174 & 0 & 5 & 4 & 0 & 31 & 35 & 100\%  \\ 
  \textbf{W60\_90D} & 1684 & 0 & 0 & 4 & 0 & 16 & 20 & 100\% \\ 
  \textbf{W90\_120D} & 1396 & 0 & 0 & 11 & 0 & 47 & 39 & 99\%  \\ 
  \textbf{PrfMatBal} & 493 & 0 & 0 & 4 & 0 & 148 & 6 & 100\%  \\ 
  \textbf{NPrfMatBal} & 999 & 0 & 0 & 136 & 0 & 1566 & 43 & 43\% \\ 
  \textbf{B120D} & 2495 & 0 & 3 & 16 & 0 & 97 & 277 & 90\% \\ 
  \hline
  Totals & 575922 & 10 & 89 & 545 & 0 & 5493 & 1710 & 3\% \\ 
   \hline
\end{tabular}
\end{table}


\begin{table}[!htbp]
\tiny
\centering
\caption{Out of sample Confusion Matrix Misclassification Error for Bankruptcy is close to 0\% during COVID-19}
\label{BK: Confusion Matrix}
\begin{tabular}{rrrrl}
  \hline
 & \textbf{Non-BK} & \textbf{BK} & Error \\ 
  \hline
\textbf{Non-BK} & 581697 & 13 & 0\%  \\ 
\textbf{BK} & 22 & 2037 & 1\%  \\ 
  Totals & 581719 & 2050 & 0\%  \\ 
   \hline
\end{tabular}
\end{table}

\begin{table}[!htbp]
\tiny
\centering
\caption{Variable Importance for Bankruptcy using DNN still have NOI higher than LTV}
\label{BK: Variable Importance}
\begin{tabular}{rlrrr}
  \hline
 variable & relative\_importance \\ 
  \hline
 bankruptcyflag & 1 \\ 
 state & 0.64  \\ 
 nonrecover & 0.56 \\ 
 prop & 0.54  \\ 
 hasballoon & 0.51 \\ 
 maturitytype & 0.51  \\ 
 appvalue & 0.49 \\ 
 age & 0.44  \\ 
 2Yr\_Tr & 0.43  \\ 
 securltv & 0.42 \\ 
 \textbf{noi} & 0.40  \\ 
 actrate & 0.40  \\ 
year\_month & 0.39 \\ 
 face & 0.39  \\ 
 10Yr\_Tr & 0.39  \\ 
 Unemployment & 0.38  \\ 
 actpmt & 0.38 \\ 
 pmtbas & 0.38  \\ 
 gdp & 0.37  \\ 
 securwac & 0.36  \\ 
 occrate & 0.35  \\ 
 age2 & 0.35  \\ 
 \textbf{ltv} & 0.33  \\ 
 time\_to\_maturity & 0.29 \\ 
   \hline
\end{tabular}
\end{table}


\begin{table}[!htbp]
\tiny
\centering
\caption{Out of sample Confusion Matrix Misclassification Error for REO is close to 0\% during COVID-19}
\label{REO: Confusion Matrix}
\begin{tabular}{rrrrl}
  \hline
 & Non-REO & REO & Error \\ 
  \hline
Non-REO & 364993 & 1084 & 0\%  \\ 
REO & 782 & 365042 & 0\%  \\ 
  Totals & 365775 & 366126 & 0\%  \\ 
   \hline
\end{tabular}
\end{table}

\begin{table}[!htbp]
\tiny
\centering
\caption{Variable Importance for REO using DNN no longer have NOI higher than LTV}
\label{REO: Variable Importance}
\begin{tabular}{rlrrr}
  \hline
 variable & relative\_importance \\ 
  \hline
 age & 1 \\ 
 state & 0.97  \\ 
 prop & 0.89 \\ 
 age2 & 0.85 \\ 
 actpmt & 0.79  \\ 
 \textbf{ltv} & 0.76  \\ 
 securltv & 0.75 \\ 
 securwac & 0.73  \\ 
 hasballoon & 0.72  \\ 
 time\_to\_maturity & 0.72  \\ 
 bankruptcyflag & 0.71  \\ 
 face & 0.67 \\ 
 pmtbas & 0.67  \\ 
 nonrecover & 0.66  \\ 
 maturitytype & 0.66  \\ 
 actrate & 0.60  \\ 
 \textbf{noi} & 0.59  \\ 
 appvalue & 0.58  \\ 
 occrate & 0.52  \\ 
 year\_month & 0.44  \\ 
 2Yr\_Tr & 0.40  \\ 
 Unemployment & 0.36  \\ 
 10Yr\_Tr & 0.26  \\ 
 gdp & 0.08  \\ 
   \hline
\end{tabular}
\end{table}


\begin{table}[!htbp]
\tiny
\centering
\caption{Out of sample Confusion Matrix Misclassification Error for Foreclosure is close to 0\% during COVID-19}
\label{Foreclosure: Confusion Matrix}
\begin{tabular}{rrrrl}
  \hline
 & Non-FCL & FCL & Error  \\ 
  \hline
Non-FCL & 364993 & 1084 & 0\%  \\ 
FCL & 782 & 365042 & 0\%  \\ 
  Totals & 365775 & 366126 & 0\%  \\ 
   \hline
\end{tabular}
\end{table}

\begin{table}[!htbp]
\tiny
\centering
\caption{Variable Importance for Foreclosure using DNN no longer have NOI higher than LTV}
\label{Foreclosure: Variable Importance}
\begin{tabular}{rlrrr}
  \hline
 variable & relative\_importance \\ 
  \hline
 age & 1  \\ 
 state & 0.97  \\ 
 prop & 0.89 \\ 
 age2 & 0.85 \\ 
 actpmt & 0.79  \\ 
 \textbf{ltv} & 0.76  \\ 
 securltv & 0.75  \\ 
 securwac & 0.73 \\ 
 hasballoon & 0.72 \\ 
 time\_to\_maturity & 0.72  \\ 
 bankruptcyflag & 0.71  \\ 
 face & 0.67  \\ 
 pmtbas & 0.67  \\ 
 nonrecover & 0.66  \\ 
 maturitytype & 0.66 \\ 
 actrate & 0.60  \\ 
 \textbf{noi} & 0.59  \\ 
 appvalue & 0.58  \\ 
 nonrecover.N & 0.54 \\ 
 occrate & 0.52  \\ 
 year\_month & 0.44  \\ 
 2Yr\_Tr & 0.40 \\ 
 Unemployment & 0.36  \\ 
 10Yr\_Tr & 0.26  \\ 
 gdp & 0.08  \\ 
   \hline
\end{tabular}
\end{table}


\begin{table}[!htbp]
\tiny
\centering
\caption{Variable Importance for W90\_120D loans using DNN no longer have NOI higher than LTV}
\label{W90120D: Variable Importance}
\begin{tabular}{rlrrr}
  \hline
  variable & relative\_importance  \\ 
  \hline
 state & 1  \\ 
age & 0.86 \\ 
 proptype & 0.75 \\ 
 bankruptcyflag & 0.75 \\ 
 actrate & 0.75  \\ 
 securltv & 0.74 \\ 
 hasballoon & 0.71  \\ 
 \textbf{ltv} & 0.71  \\ 
 actpmt & 0.69  \\ 
 pmtbas & 0.68  \\ 
 securwac & 0.68  \\ 
 face & 0.64  \\ 
 nonrecover & 0.64  \\ 
 \textbf{noi} & 0.62 \\ 
 age2 & 0.61  \\ 
 appvalue & 0.61  \\ 
 2Yr\_Tr & 0.60  \\ 
 maturitytype. & 0.59  \\ 
 time\_to\_maturity & 0.58  \\ 
 occrate & 0.54  \\ 
 year\_month & 0.48 \\ 
 10Yr\_Tr & 0.33  \\ 
 Unemployment & 0.31  \\ 
 gdp & 0.16  \\ 
   \hline
\end{tabular}
\end{table}

\begin{table}[!htbp]
\tiny
\centering
\caption{Out of sample Confusion Matrix Misclassification Error for W90\_120D is high during COVID-19}
\label{W90120D: Confusion Matrix}
\begin{tabular}{rrrrl}
  \hline
 & Non-W90\_120D & W90\_120D) & Error  \\ 
  \hline
Non-W90\_120D & 577640 & 4636 & 1\% \\ 
W90\_120D & 1365 & 128 & 91\%  \\ 
  Totals & 579005 & 4764 & 1\%  \\ 
   \hline
\end{tabular}
\end{table}


\begin{table}[!htbp]
\tiny
\centering
\caption{Variable Importance for B120D loans using DNN no longer have NOI higher than LTV}
\label{B120D: Variable Importance}
\begin{tabular}{rlrrr}
  \hline
  variable & relative\_importance  \\ 
  \hline
 age & 1  \\ 
  state & 0.97  \\ 
   prop & 0.89 \\ 
   age2 & 0.85  \\ 
   actpmt & 0.79 \\ 
  \textbf{ltv} & 0.76  \\ 
   securltv & 0.75  \\ 
   securwac & 0.73  \\ 
  hasballoon & 0.72  \\ 
   time\_to\_maturity & 0.72  \\ 
   bankruptcyflag & 0.71 \\ 
   face & 0.67 \\ 
   pmtbas & 0.67  \\ 
  nonrecover & 0.66  \\ 
   maturitytype & 0.66  \\ 
   actrate & 0.60  \\ 
   \textbf{noi} & 0.59  \\ 
   appvalue & 0.58  \\ 
   occrate & 0.52  \\ 
   year\_month & 0.44  \\ 
   2Yr\_Tr & 0.40 \\ 
   Unemployment & 0.36  \\ 
   10Yr\_Tr & 0.26  \\ 
   gdp & 0.08  \\ 
   \hline
\end{tabular}
\end{table}

\begin{table}[!htbp]
\tiny
\centering
\caption{Misclassification Error is high for B120D loans}
\label{B120D: Confusion Matrix}
\begin{tabular}{rrrrl}
  \hline
 & Non-B120D & B120D & Error \\ 
  \hline
Non-B120D & 579326 & 1555 & 0\% \\ 
B120D & 2382 & 506 & 82\%  \\ 
  Totals & 581708 & 2061 & 1\% \\ 
   \hline
\end{tabular}
\end{table}

\end{appendices}
\end{document}